\documentclass{aastex}
\newcommand{\hm}{H$_{2}$}

\newcommand{\nhp}{H$_{P}$}
\newcommand{\nhc}{H$_{C}$}

\begin{document}

\title{Molecular hydrogen formation on dust grains in the high
redshift universe.}  
 \author{S.~Cazaux and M.~Spaans } \affil{Kapteyn
 Astronomical Institute} \affil{P.O.~Box 800, NL--9700 AV Groningen,
 The~Netherlands} \email{Stephanie Cazaux (cazaux@astro.rug.nl)}
 \email{Marco Spaans (spaans@astro.rug.nl)}

\begin{abstract}

We study the formation of molecular hydrogen on dust grain surfaces
and apply our results to the high redshift universe. We find that a
range of physical parameters, in particular dust temperature and gas
temperature, but not so much dust surface composition, influence the
formation rate of H$_2$.  The H$_2$ formation rate is found to be
suppressed above gas kinetic temperatures of a few hundred K and for
dust temperatures above 200-300 K and below 10 K. We highlight the
differences between our treatment of the H$_2$ formation process and
other descriptions in the literature.  We also study the relative
importance of H$_2$ formation on dust grains with respect to molecular
hydrogen formation in the gas phase, through the H$^-$ route. The
ratio of formation rates of these two routes depends to a large part
on the dust abundance, on the electron abundance, and also on the
relative strength of the FUV (extra-)galactic radiation field.  We
find that for a cosmological evolution of the star formation rate and
dust density consistent with the Madau plot, a positive feedback
effect on the abundance of H$_2$ due to the presence of dust grains
can occur at redshifts $z\ge 3$. This effect occurs
for a dust-to-gas mass ratio as small as 10$^{-3}$ of the galactic
value.

\end{abstract} 

\section{Introduction}

\hm\ formation in the universe is a process which has been studied
extensively in the past decades, but which is still not well
understood. In the interstellar medium, \hm\ formation occurs on grain
surfaces, permitting three-body reactions that are much more efficient
than gas phase reactions (Gould \& Salpeter 1963). Many studies have
been performed over the years and focussed on explaining the
mechanisms involved in the \hm\ formation process on grain
surfaces. It is clear that this process is governed by the mobility of
the H atoms on the grain, as well as the energies binding the H atoms
to the surface. The mobility of an adatom (the atom that is bound to
the surface) is the combination of two processes: thermal diffusion
and quantum tunneling. Therefore, an adatom can move on the surface of
the grain, from site to site, through these two processes according to
the temperature of the grain and the characteristics of the surface
(Barlow \& Silk 1976; Leitch-Devlin \& Williams 1984; Tielens \&
Alamandola 1987). On astrophysically relevant surfaces (olivine and
amorphous carbon), an adatom can bind in two energetically different
sites: physisorbed sites and chemisorbed sites. Physisorbed atoms are
weakly bound to the surface and are mobile for low grain temperatures
(around 10 K), whereas chemisorbed atoms are bound stronger to the
surface and become mobile for a grain temperature of a few hundred K
(Barlow \& Silk 1976; Aronowitch \& Chang 1980; Klose 1992; Fromherz
et al.\ 1993; Cazaux \& Tielens 2002, 2004). Therefore, \hm\ formation
strongly depends on the metallicity of the system (i.e., the dust
abundance), the considered grains (olivine or amorphous carbon), the
temperature of the grains and the atomic hydrogen abundance.

Studies of \hm\ formation in low metallicity systems -- e.g., the high
redshift universe -- have already been performed (c.f.\ Tegmark et
al.\ 1997; Glover 2003).  \hm\ at high redshift is of prime importance
since it is considered to be the only coolant below $10^4$
K. Therefore, as discussed by many authors, \hm\ plays a crucial role
in the formation of the first stars (Haiman, Rees, \& Loeb 1996;
Haiman, Thoul, \& Loeb 1996; Tegmark et al.\ 1997), and thus
determines the end of the dark ages (Haiman, Abel, \& Rees
2000). Other authors discussed the possibility of HD as a more
efficient coolant because this molecule possesses a nonzero dipole
moment, even though its abundance is much lower (Galli \& Palla 1998;
Nakamura \& Umemura 2002). Norman \& Spaans (1997) estimate a redshift
of roughly unity, where \hm\ formation in the gas phase and \hm\
formation on grain surfaces are equally important. At high redshift,
\hm\ formation proceeds through the H$^-$ route.  Below a redshift of
unity, grain surface reactions become more important relative to the
(inefficient) H$^-$ route. Note that at high densities H$_2$ is formed
through three-body association.

These issues are of more than academic interest since the presence of
H$_2$, either through the H$^-$ or dust grain routes, is instrumental
in the ability of low metallicity gas to cool and contract to form
stars. Because no dust grains are available prior to the formation of
pop III objects, the H$^-$ route is crucial in starting up the star
formation process, but the subsequent expulsion of metals in supernova
explosions and ambient dust formation can strongly enhance the
formation of molecular hydrogen (Hirashita, Hunt, \& Ferrara 2002).
\\The construction of comprehensive models for the effects of dust
grains on early galaxy evolution and their observational consequences
is underway (c.f.\ Hirashita, Hunt, \& Ferrara 2002; Spaans \& Norman
1997; Kauffmann \& Charlot 1998; Somerville, Primack \& Faber 2001;
Glover 2003), motivated by upcoming observatories like ALMA and
JWST. In general, such studies adopt the H$_2$ formation rate of
Hollenbach \& McKee (1979). Although a good order of magnitude
estimate, recent work (Cazaux \& Tielens 2002, 2004) has allowed the
formation of H$_2$ (and HD) on dust grains to be computed more
accurately.  Because the H$_2$ chemistry plays such an important role
in the high redshift universe it is crucial to put it on the firmest
of physical bases.  The aim of this paper is then to provide detailed
computations of the H$_2$ formation efficiency and formation rate
as functions of ambient physical conditions and to extend earlier work
on the effects of dust grains on the physical properties of primordial
gas.

In this paper, we use the model developed by Cazaux \& Tielens (2004)
and extend it to accommodate the conditions in the high redshift
universe.  This model, which describes the formation of \hm\ on grain
surfaces, takes into account the presence of both physisorbed and
chemisorbed sites on the grain surface, and allows quantum mechanical
diffusion as well as thermal diffusion for adatoms to go from site to
site. In Section two, we calculate first the \hm\ formation
efficiency for typical H fluxes, different surface characteristics,
and different dust grain abundances, and second the \hm\ formation
efficiency through the gas phase route (i.e., the H$^-$ route). In
Section three, we discuss which are the physical conditions (i.e.,
dust and gas temperature, dust abundance, electron fraction) at which
the H$^-$ route and grain surface route make equal contributions to
the \hm\ formation rate. Then, we adopt a cosmological model in order
to assess at which redshift this equality occurs, consistent with the
star formation rate history described by Madau ``the Madau
plot'' (e.g., the universal star formation rate as a function of
redshift; Lilly et al.\ 1996; Madau et al.\ 1996). Therefore, we
estimate when, in the history of the universe, the dust route became
the dominant process toward \hm\ formation. In Section four, we
discuss the effects of this gas to dust route transition on the star
formation rate and the ambient radiation field.

\section{\hm\ Formation}

\subsection{Dust grain Route}
In this section we study the formation of \hm\ on interstellar grain
surfaces. First, we discuss the typical range of the H flux in diffuse
interstellar clouds. Second, we study the intrinsic properties of the
grain (olivine or carbonaceous), the impact of the parameters
characterizing the surface of the grain, and the role of the grain
size distribution.

\subsubsection{H Flux}

A grain in the ISM is irradiated by H atoms from the gas
phase. The H flux in a diffuse interstellar cloud is given by
\begin{equation}
F_{\rm H}=\frac{n({\rm H})v_{\rm H}}{N_S},
\end{equation}
where $N_S$ is the number of sites per $cm^2$ on the surface of the
grain, $n({\rm H})$ the density of H atoms in the gas phase,
and $v_{\rm H}$ the mean
velocity of these atoms.  We assume $N_S=2\times 10^{15}$ sites
cm$^{-2}$ for a 0.1 $\mu$m grain, the density of H atoms to be between
1 and 100 particles per cm$^3$, and the velocity of these atoms to be
between 1 and 10 km s$^{-1}$.\\ Therefore, we consider a range for
the flux of $5\times 10^{-11}\le F_{\rm H} \le 5\times 10 ^{-8}$ where
$F_{\rm H}$ is the flux of H atoms in monolayer per second (mLy s$^{-1}$).

\subsubsection{Characteristics and nature of the Grains}

\begin{figure}
\plotone{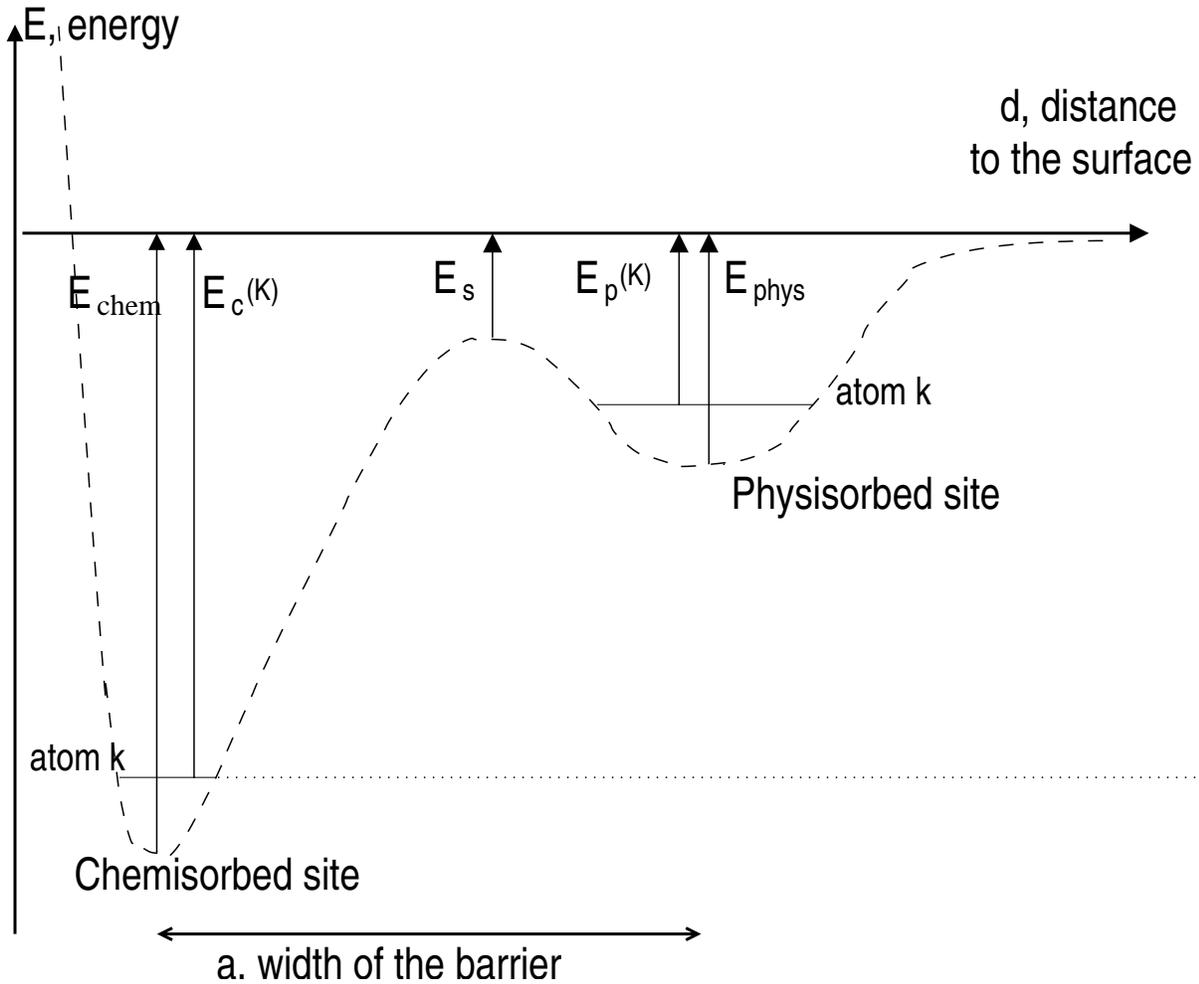}
\caption{Barrier between a physisorbed site and a chemisorbed site for
an atom, $k$, bound to the surface. When an atom crosses this barrier,
it moves perpendicularly to the surface. Here, we consider hydrogen or
deuterium atoms. Their energies are different because D atoms are more
tightly bound to the chemisorbed and physisorbed sites than H atoms
due to zero point energy difference. $E_{s}$ is the energy of the
saddle point. In this work only H atoms are considered.}
\label{barrier}
\end{figure}

\subsubsubsection{Surface Characteristics} In this section we present
the characteristics of the grain surface that are relevant to the
model we use in this paper.  This model has been discussed by
Cazaux \& Tielens (2004), and describes how \hm\ forms on grain
surfaces. \\ When an atom hits a grain, it can either be bound to the
surface, if it arrives in an empty site, or it can go back into the
gas phase, if the site is occupied. This process follows Langmuir
kinetics. An adatom can bind to the surface in two energetically
different sites: a chemisorbed site or a physisorbed site. According
to the interaction between the atom and the surface, the mobility of
the adatom is set. Coming from the gas phase the atoms are first
physisorbed. Then they can either cross the barrier to go to a
chemisorbed site (depicted in Fig.\ ~\ref{barrier}) by moving
perpendicular to the surface, or go to another physisorbed site, by
moving along the surface. Considering an adatom H, the relevant
parameters for our study are the desorption energies of chemisorbed
hydrogen, $E_{{\rm H}_C}$, of physisorbed hydrogen $E_{{\rm H}_P}$,
and of molecular hydrogen $E_{\rm H_2 }$, as well as the energy of the
saddle point between a physisorbed and a chemsiorbed site $E_S$ and
the factor $\mu$ which is the fraction of the newly formed molecules
which stays on the surface. Fig.\ ~\ref{barrier} illustrates these
parameters, in the case of a barrier between a physisorbed and a
chemisorbed site. Under steady state conditions, the \hm\
formation efficiency (i.e., the fraction of incoming atoms leaving
the grain as \hm\ molecules) varies with these parameters. When we
know the different physical processes involved in \hm\ formation on
grain surfaces, we can understand the impact of each parameter on the
\hm\ formation, and at which range of gas and dust temperatures it
occurs.

\begin{figure}
\plottwo{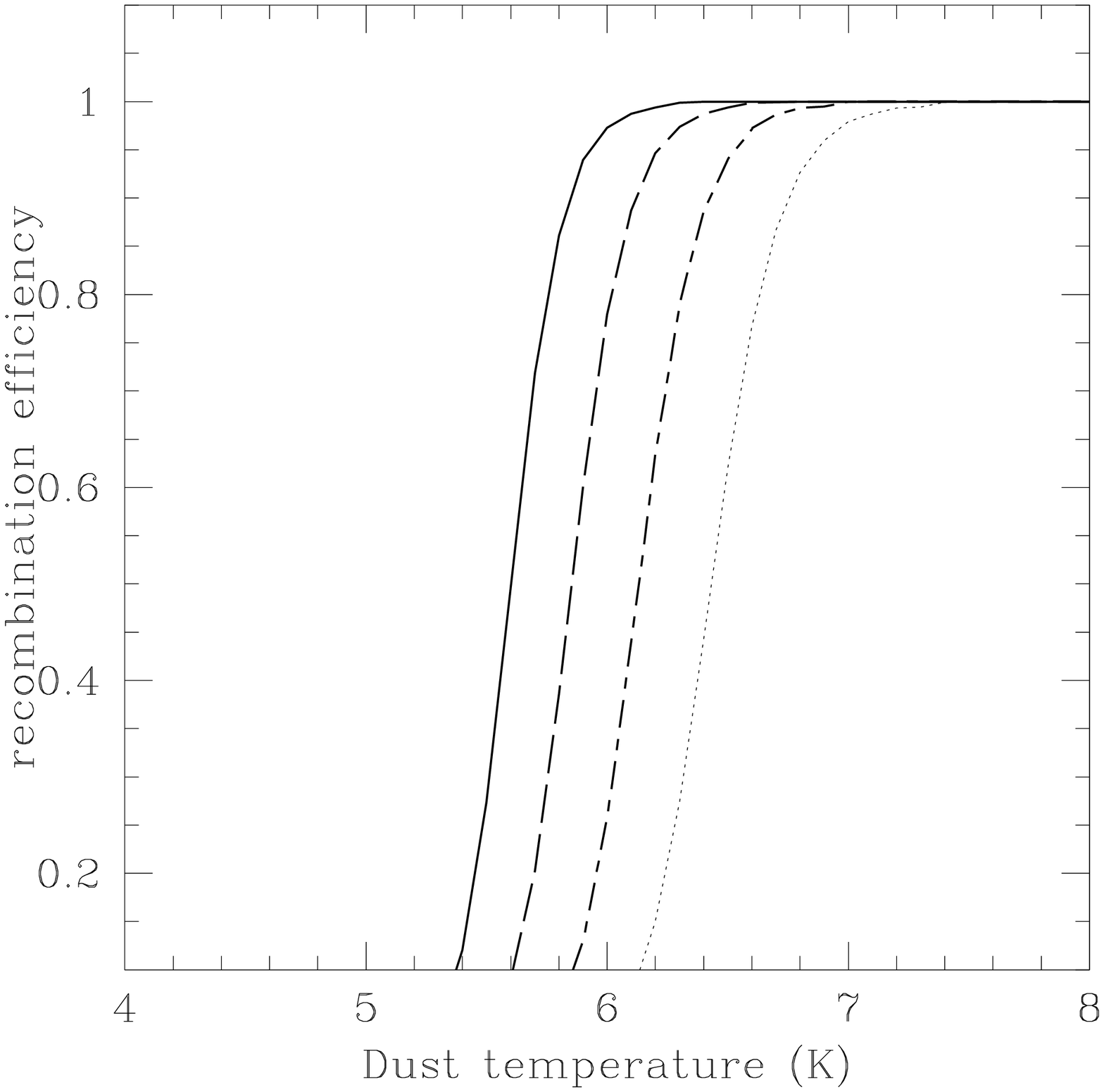}{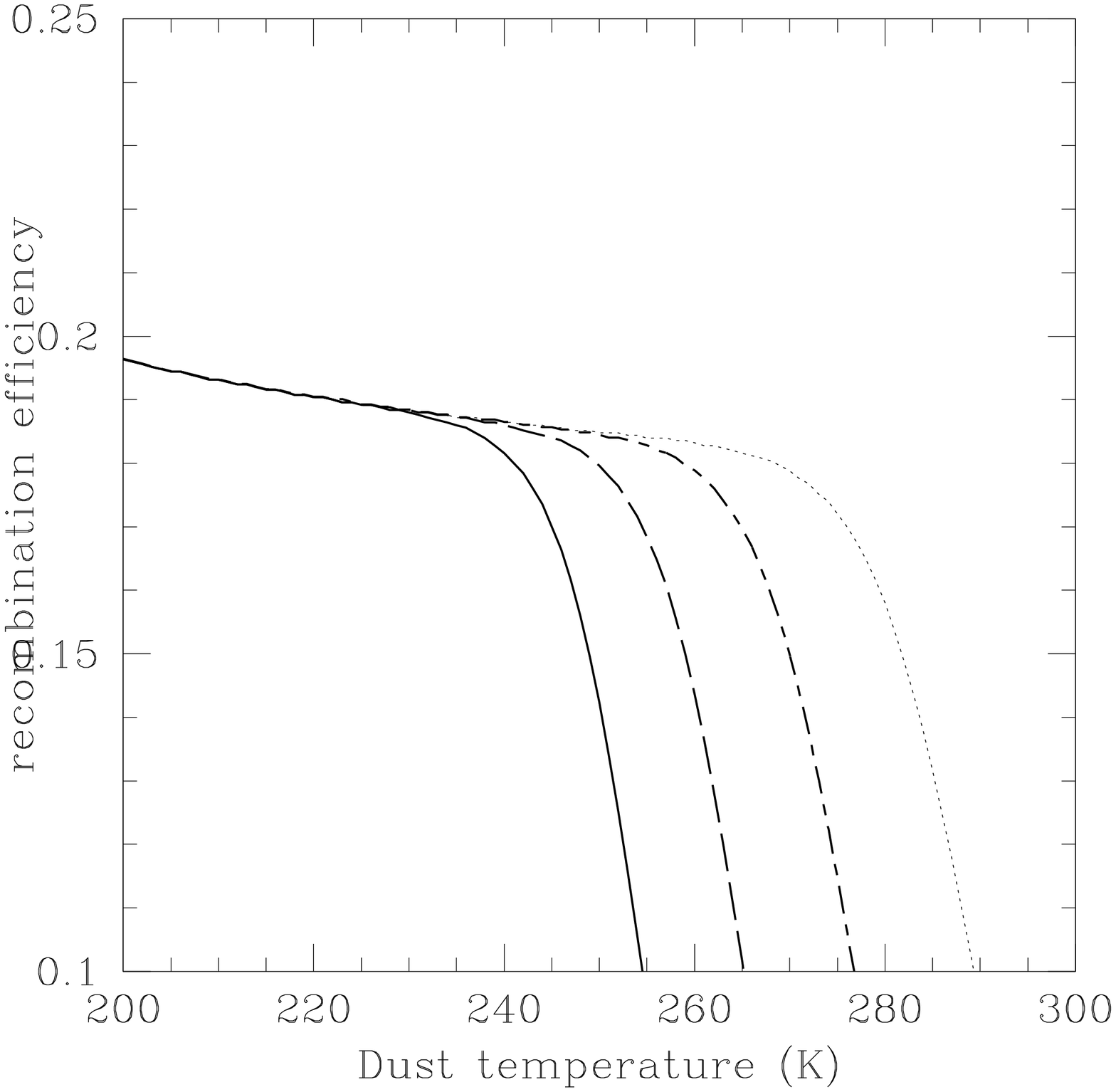}
\caption{\hm\ formation efficiency at low and high dust temperatures
for 4 different fluxes: solid line $5\times 10^{-11}$, dashed line
$5\times 10^{-10}$,
dot-dashed $5\times 10^{-9}$, dots $5\times 10^{-8}$. The parameters are
$E_{\rm H_2}=340$ K, $E_{{\rm H}_P}=600$ K, $E_{S}=200$ K,
$E_{{\rm H}_C}=10,000$ K and $\mu =0.4$.}
\label{st}
\end{figure}

At low and high dust temperatures, variations in the flux affect the
\hm\ formation efficiency. At low dust temperatures, this can be
explained by the Langmuir kinetics. If the flux is low, the atoms,
after a stay in the physisorbed sites, move into the chemisorbed sites
without encountering other incoming atoms from the gas phase. If the
flux is large, the physisorbed atoms, before moving to the chemisorbed
sites, encounter some incoming atoms from the gas phase and are both
released into the gas phase again. This explains the less efficient
\hm\ formation at high fluxes. On the contrary, at high dust
temperatures, only a small fraction of the atoms coming onto the grain
go to chemisorbed sites. This fraction is set by the time required to
evaporate from a physisorbed site, or to move to a chemisorbed
site. When the flux increases, the number of chemisorbed H atoms just
increases too, and the \hm\ formation is more efficient since the
grain surface is occupied by more H atoms. This impact of the flux on
the \hm\ formation efficiency is presented Fig. ~\ref{st}.

\begin{figure}
\plottwo{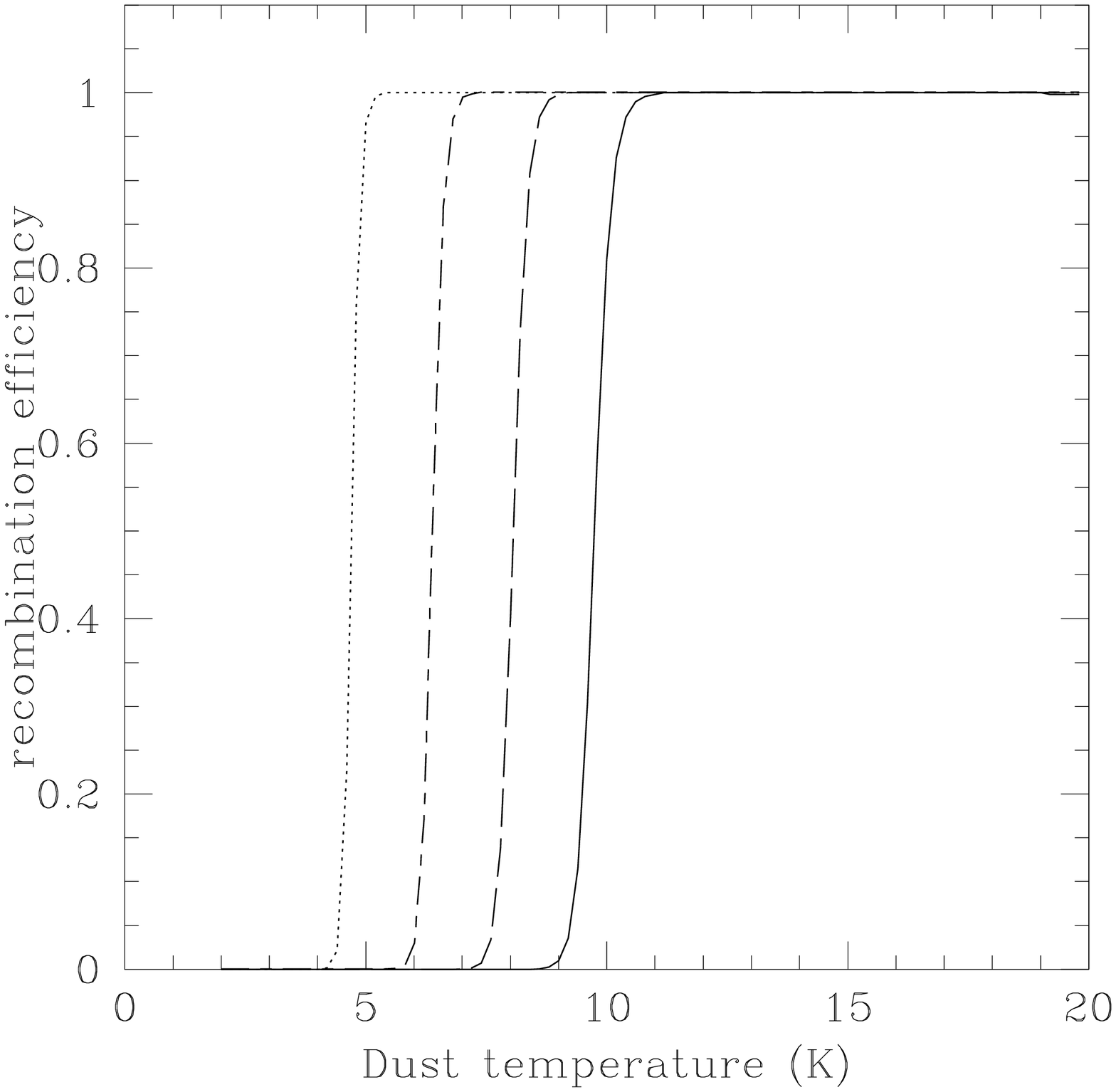}{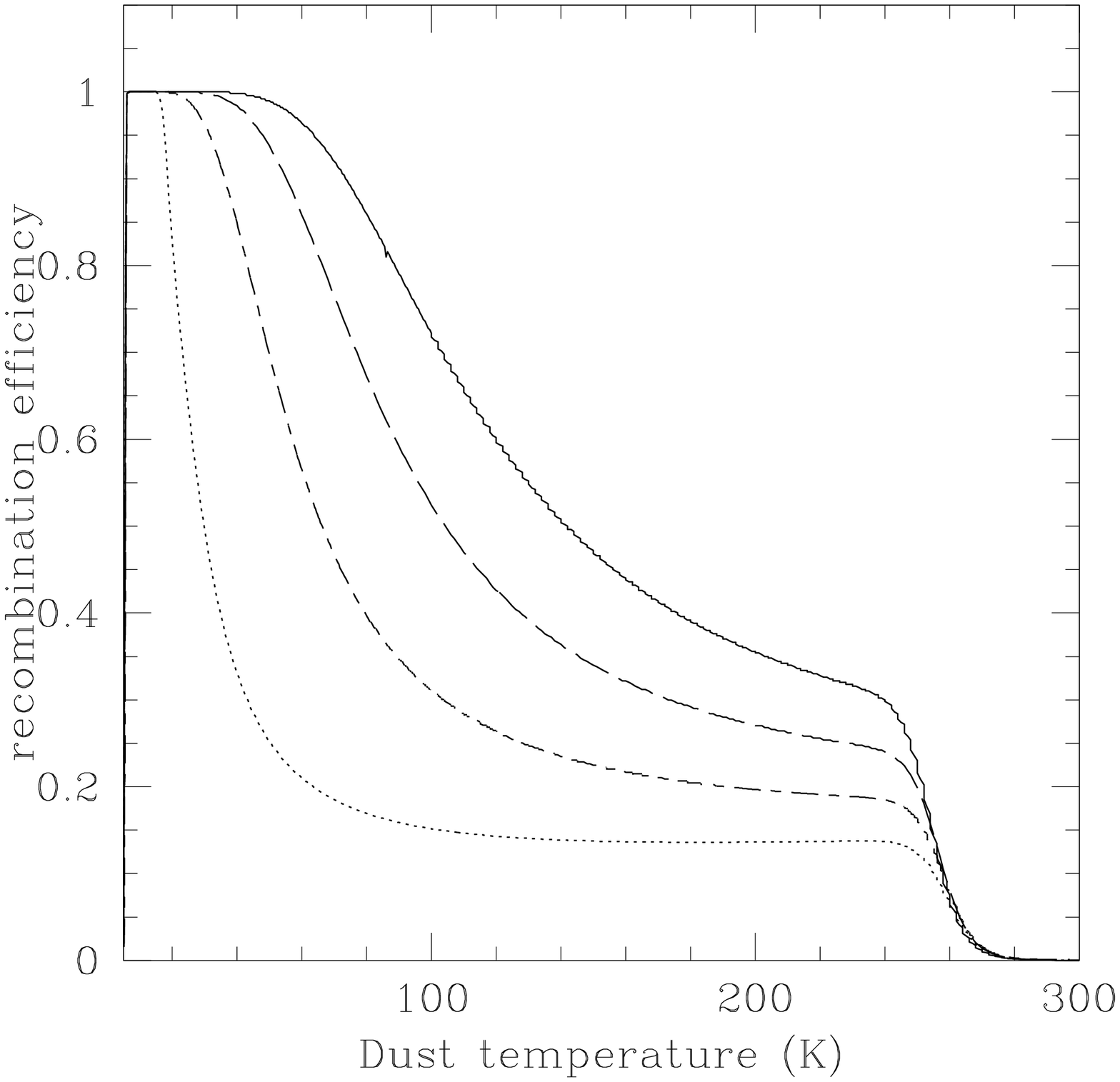}
\caption{Left.\ \hm\ formation efficiency for 4 different energies of
\hm\ desorption, $E_{{\rm H}_2}$: Solid line 600 K, dashed line 500 K,
dot-dashed 400 K, dots 300 K. Fixed parameters are $F_{\rm
H}=10^{-10}$, $E_{{\rm H}_P}=600$ K, $E_{S}=200$ K, $E_{{\rm
H}_C}=10,000$ K. Right.\ \hm\ formation efficiency for 4 different
energies of the saddle point E$_S$: Solid line 400 K, dashed line 300
K, dot-dashed 200 K, dots 100 K. Fixed parameters are $F_{\rm
H}=10^{-10}$, $E_{{\rm H}_P}=600$ K, $E_{\rm H_2}=340$ K, $E_{{\rm
H}_C}=10,000$ K.}
\label{E0}
\end{figure}

Another parameter which affects the \hm\ formation at low dust
temperatures is the \hm\ desorption energy. \hm\ can form and stay on
the surface of the grain until a temperature is reached that allows
evaporation of these molecules. This desorption process is driven by
the desorption energy of \hm , $E_{\rm H_2 }$, and consequently this
parameter has a big impact on the \hm\ formation efficiency, as
presented in fig~\ref{E0}, left panel.

\begin{figure}
\plottwo{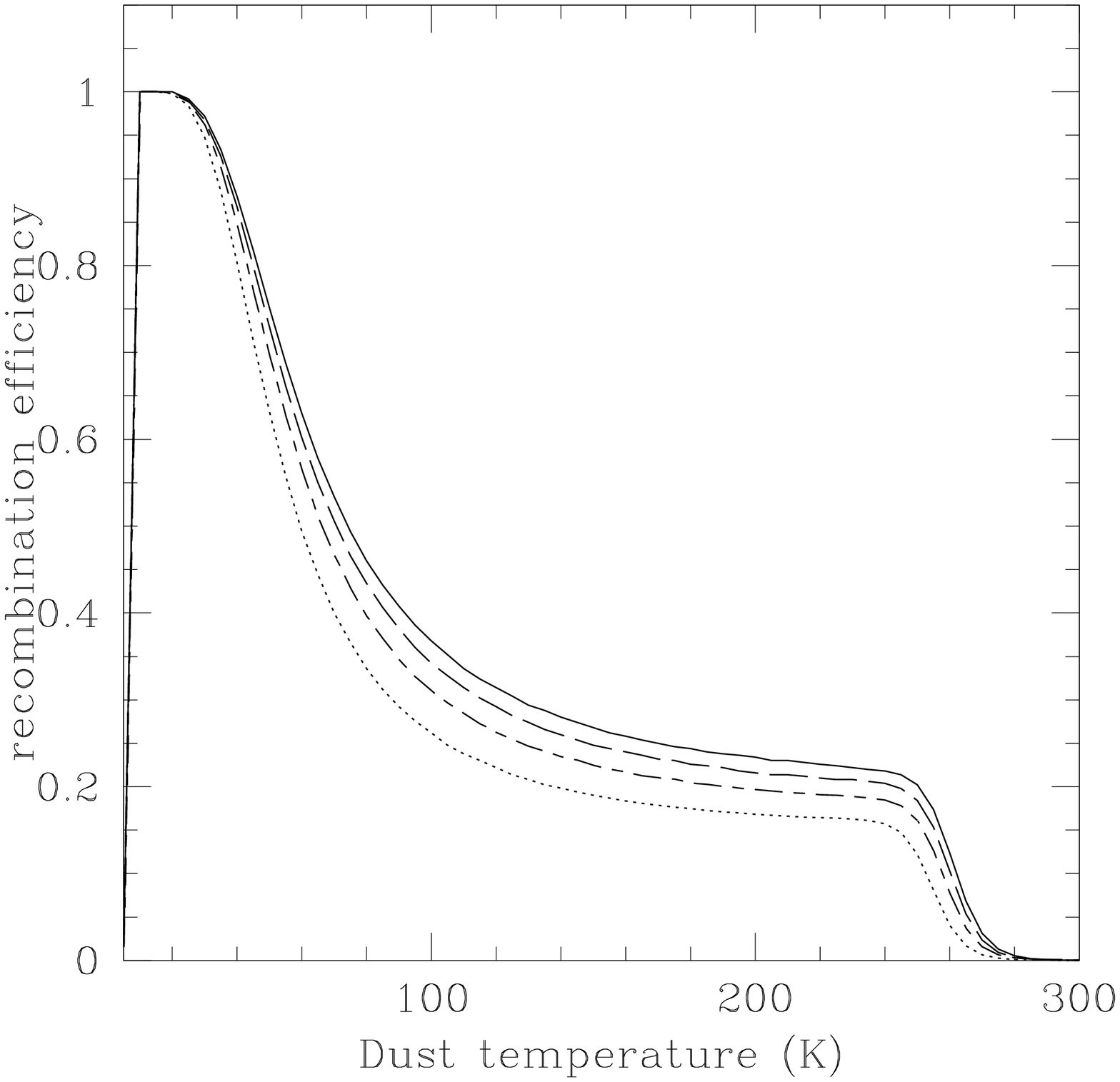}{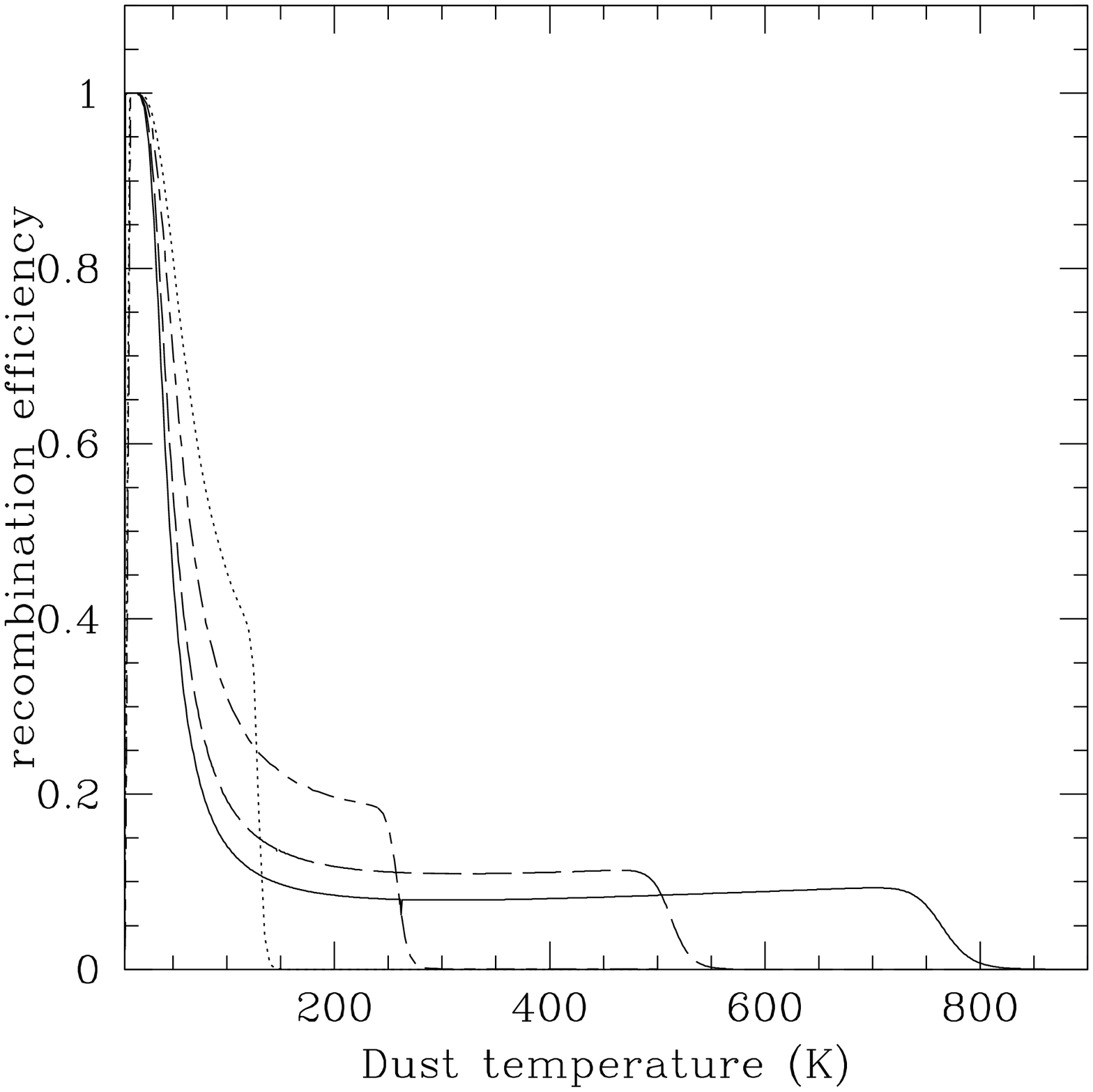}
\caption{Left.\ \hm\ formation efficiency for 4 different desorption
energies of physisorbed H, $E_{{\rm H}_C}$: solid line 400 K, dashed
line 600 K, dot-dashed 800 K, dots 1000 K. Fixed parameters are
$F_{\rm H}=10^{-10}$, $E_{\rm H_2}=340$ K, $E_{S}=200$ K, $E_{{\rm
H}_C}=10,000$ K. Right.\ \hm\ formation efficiency for 4 different
energies of chemisorbed H desorption, $E_{{\rm H}_C}$: Solid line
30,000 K, dashed line 20,000K, dot-dashed 10,000 K, dots 5000 K. Fixed
parameters are $F_{\rm H}=10^{-10}$, $E_{\rm H_2}=340$ K, $E_{{\rm
H}_P}=600$ K, $E_{S}=200$ K.}
\label{Epc}
\end{figure}

At higher dust temperatures, the desorption energy of physisorbed
atoms, as well as the energy of the saddle point, affect the \hm\
formation process. Indeed, the only obstacle to form \hm\ at higher
dust temperatures, is the evaporation of the physisorbed H atoms
before association. At these dust temperatures the physical process
to form \hm\ is the encounter of a physisorbed H and a chemisorbed H
atom. This means that the incoming H atom, which is physisorbed, has
to cross the barrier presented in Fig.\ ~\ref{barrier}, and hence the
energies $E_{{\rm H}_P}$ and $E_S$ have a big impact on the \hm\
formation efficiency, as shown in Fig.\ ~\ref{E0}, right panel and
Fig.\ ~\ref{Epc}, left panel.

Finally, when the dust temperature is too high to enable the
physisorbed atoms to recombine before they evaporate, the \hm\
formation process reduces to the association of two chemisorbed H
atoms. Of course, this process depends on the desorption energy of the
chemisorbed atoms, as illustrated in Fig.~\ref{Epc}, right panel. When
the dust temperature is high enough to enable chemisorbed H
evaporation, these atoms leave the grain before recombining, and \hm\
formation is quenched.

\subsubsubsection{Olivine versus Carbon} 

\begin{figure}
\includegraphics[width=16cm,keepaspectratio,angle=-90]{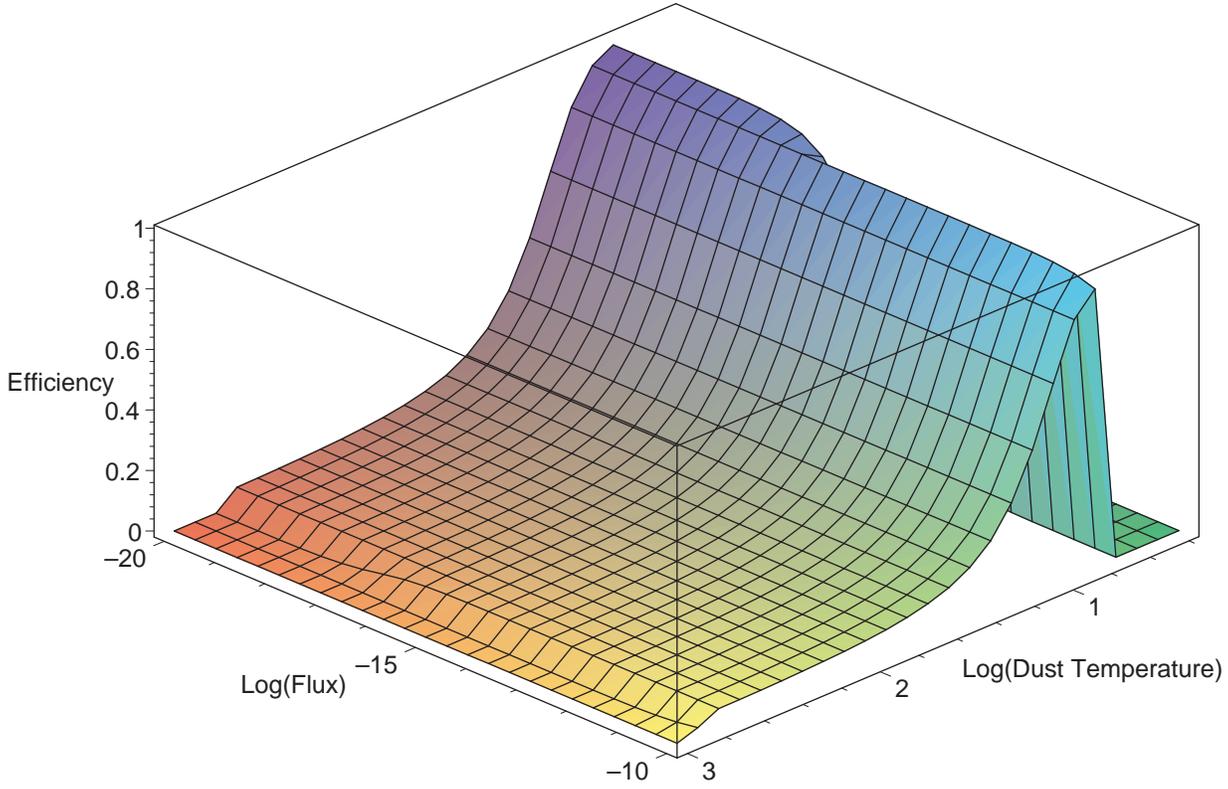}
\caption{\hm\ formation efficiency as a function of the dust temperature
and the flux for olivine grains.}
\label{FT}
\end{figure}

\begin{figure}
\includegraphics[width=16cm,keepaspectratio,angle=-90]{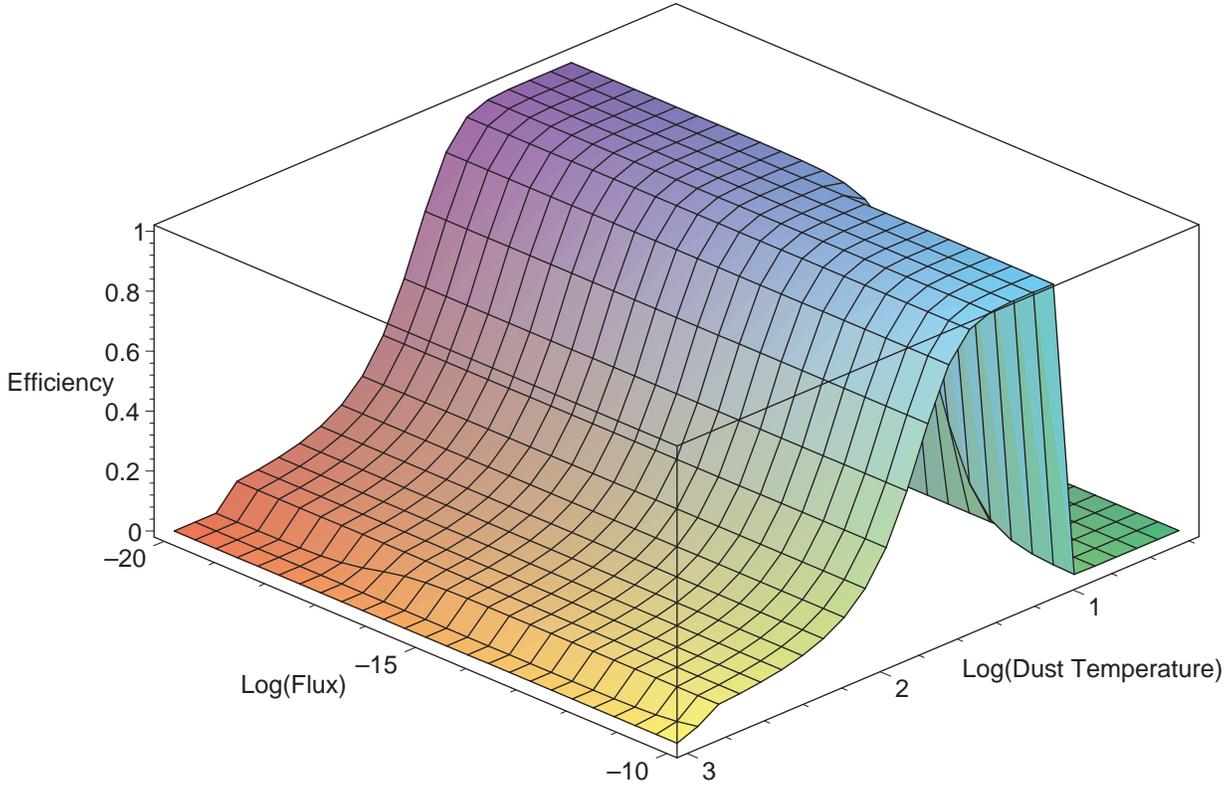}
\caption{\hm\ formation efficiency as a function of the dust temperature
and the flux for amorphous carbon grains.}
\label{FTc}
\end{figure}

The composition of dust in the diffuse interstellar medium is still
uncertain.  According to observations, this composition includes
silicates, amorphous carbon, polycyclic aromatic hydrocarbons (PAHs),
graphite organic refractories, and many more (Mathis, Rumpl, \&
Nordsieck 1977).  Most models combine all these elements to obtain the
interstellar extinction curve and to compare it with
observations. Weingartner \& Draine (2001) and Li \& Draine (2001,
2002) have developed a carbonaceous-silicate grain model which
successfully reproduces observed interstellar extinction, scattering
and infrared emission. This model consists of a mixture of
carbonaceous and silicate grains with a grain size distribution chosen
to reproduce the extinction curves obtained by observing the Milky
Way, the Large Magellanic Cloud (LMC) and the Small Magellanic
Cloud. This model requires the presence of very small carbonaceous
grains, and appears to be a viable explanation for the observations.
Therefore, we calculate \hm\ formation only on carbon and silicate
grains. The surface characteristics of these two grains have been
discussed in Cazaux \& Tielens (2004). The experiments done by
Pirronello et al.\ (1997a, 1997b and 1999) and Katz et al.\ (1999)
have benchmarked the model that we are using here (Cazaux \& Tielens
2004), and the derived parameters are reported in table~\ref{table1}.
The \hm\ formation efficiency on olivine and carbonaceous grains
as functions of the flux and the temperature is presented in Fig.\
~\ref{FT} and ~\ref{FTc}.

\subsubsubsection{The \hm\ formation rate}

\begin{figure}
\includegraphics[width=16cm,keepaspectratio,angle=-90]{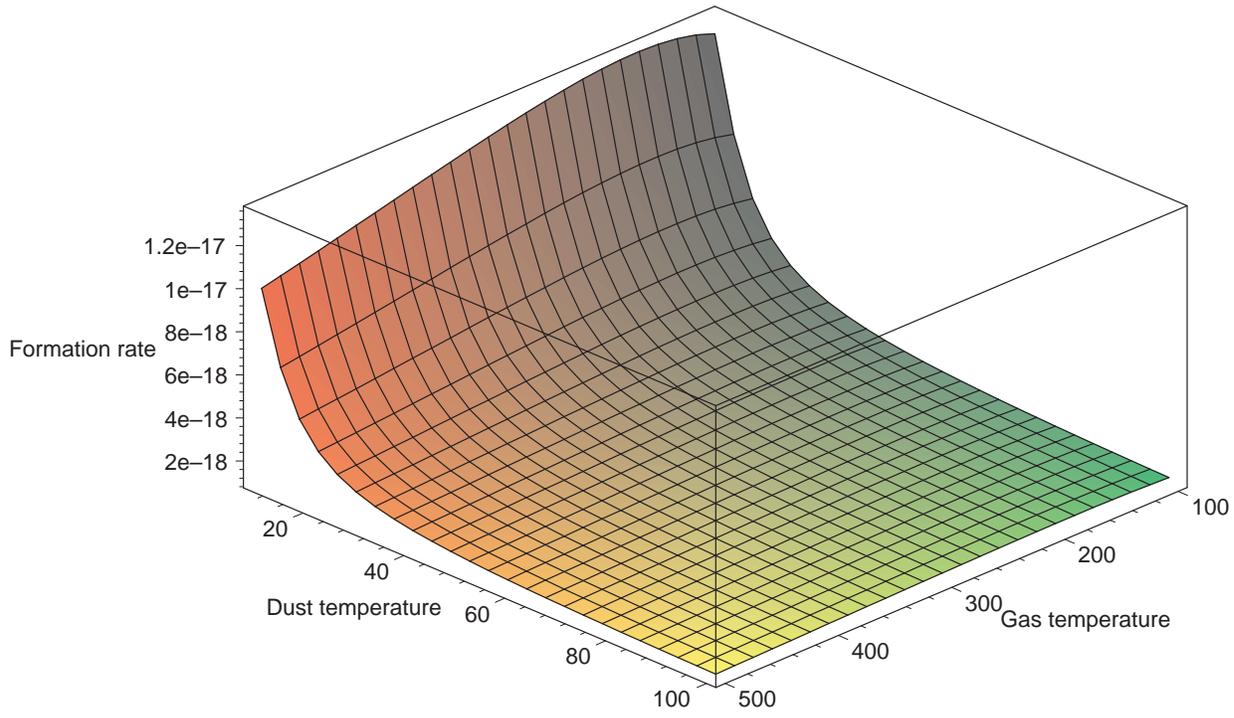}
\caption{\hm\ formation rate as a function of the dust and the gas
temperature for olivine grains.}
\label{Rolivine}
\end{figure}

\begin{figure}
\includegraphics[width=16cm,keepaspectratio,angle=-90]{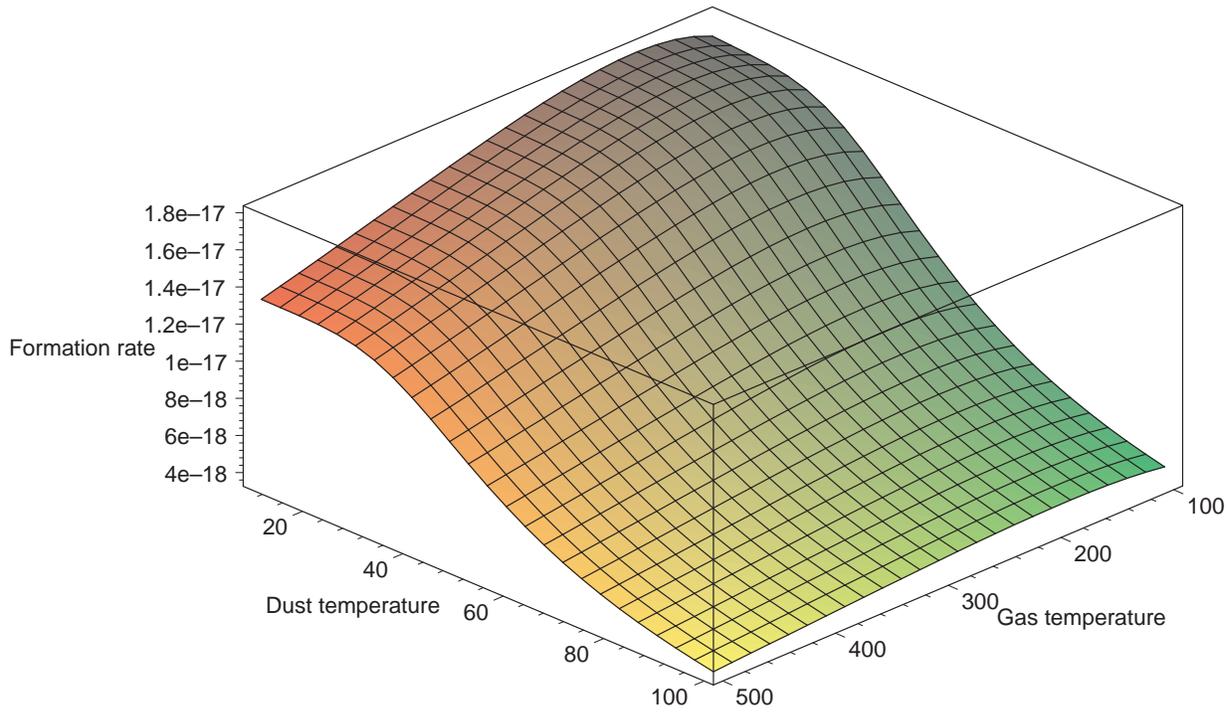}
\caption{\hm\ formation rate as a function of the dust and the gas
temperature for amorphous carbon grains.}
\label{Rcarbon}
\end{figure}

In astrophysical environments, the formation rate can be written
as
\begin{equation}
R_d={\frac{1}{2}} n({\rm H}) v_{\rm H} n_d \sigma_d \epsilon_{\rm H_2 } S_{\rm H},
\end{equation}
where $n({\rm H})$ and $v_{\rm H}$ are the number density and the
thermal velocity of $H$ atoms in the gas phase, $n_d \sigma_d$ is the
total cross section of interstellar grains, $\epsilon_{\rm H_2 }$ is
the formation efficiency that is discussed in detail in (Cazaux \&
Tielens 2004) and $S_{\rm H}$ is the sticking coefficient of the H
atoms which depends both on the dust and the gas temperature. $S_{\rm
H}$ is given by (Hollenbach and McKee 1979)
\begin{equation} 
S_{\rm H}(T)= \left(
1+0.4\times\left(\frac{T_g+T_d}{100}\right)^{0.5}+0.2\times\frac{T_g}{100}+0.08\times\left(\frac{T_g}{100}\right)^2\right)^{-1},
\end{equation}
where $v_{\rm H}$ is of the order of $1.45\times 10^5
\sqrt{\frac{T_g}{100}}$cm s$^{-1}$. We consider typical grains with a
radius of 0.1 $\mu$m and a material density of 3 g cm$^{-3}$
(Hollenbach and McKee 1979). Therefore, the typical mass of a dust
grain is about $1.256 \times 10^{-14}$ g. We can write
$n_d=1.329\times 10^{-10} \xi_d n_{\rm H}$, where $\xi_d$ is the dust
to gas mass ratio, which is equal to 0.01 under Galactic conditions,
and $n_{\rm H}$ is the density of hydrogen in all forms. The
formation rate can be written as
\begin{equation}
R_d=3.025\times 10^{-17}\epsilon_{\rm H_2 } \frac{\xi_d}{0.01} n_{\rm H} n({\rm H}) S_{\rm H} \sqrt{\frac{T_g}{100}} cm^{-3} s^{-1},
\end{equation}
as previously calculated by Tielens \& Hollenbach (1985).  Due to the
functional dependence of $S_{\rm H}$ on the gas and dust temperature,
the choice of olivine or amorphous carbon as the substrate is of
little consequence for the temperature behavior of $R_d$ and we take
olivine as the substrate in the remainder of the paper. This choice is
illustrated in Fig.~\ref{Rolivine} and ~\ref{Rcarbon}, where $R_d$ is
plotted as a function of $T_g$ and $T_d$ for a hydrogen density 1
cm$^{-3}$ and $\xi_d=0.01$. For the gas and dust temperatures of
interest in this work, the substrates give similar behaviors for $R_d$
within a factor of two.

\subsection{The H$^-$ Route}

In the absence of dust particles, H$_2$ can be formed through gas
phase reactions.  This gas phase route is driven by the association
of H atoms with H$^-$ ions,
\begin{equation}
{\rm H}+{\rm H}^{-}\rightarrow {\rm H}_{2}+e^{-}, 
\end{equation}
and has a rate coefficient $k_2$ that is approximately constant with
temperature below 300 K, with $k_2=1.5\times 10^{-9}$ cm$^3$ s$^{-1}$,
and equals $k_2=4.0\times 10^{-9}$ T$^{-0.17}$cm$^3$ s$^{-1}$ above
300 K (Glover 2003; Launay, Le Dourneuf \& Zeippen
1991). The equilibrium H$^{-}$ density is determined by the relation
(Donahue \& Shull 1991; De Jong 1972)
\begin{equation}
n_{\rm H^{-}}=\frac{k_1 n_e n_{\rm H}}{k_2 n_{\rm H} + k_3 J_{21} + k_4 n_{\rm H^+}}, 
\end{equation}
where $n({\rm H})\approx n_{\rm H}$ and k$_1$ is the formation rate of
H$^{-}$ via electron attachment, ${\rm H}+e^{-}\rightarrow {\rm
H}^{-}+h\nu$, with $k_1=1.4\times
10^{-18}{T_g}^{0.928}\exp{(-\frac{Tg}{16200})}$. We consider three
main processes responsible for the disappearance of H$^{-}$. First,
associative detachment, ${\rm H}^{-}+{\rm H} \rightarrow {\rm
H}_2+e^-$, that we consider to be the most important gas phase route
to form H$_2$, with a rate coefficient $k_2$ (as discussed
above). Second, photo-detachment, ${\rm H}^-+h\nu\rightarrow {\rm
H}+e^-$ with a rate coefficient, as discussed by Glover (2003) and De
Jong (1972), given by $k_3=4\pi \int_{\nu_{th}}^{\infty}
\frac{\sigma_\nu J_\nu}{h\nu} d\nu$, where $\sigma_\nu$ is the
photo-detachment cross section of H$^-$, J$_\nu$ is the intensity of
the radiation field, and $\nu_{th}$ is the energy for the
photo-detachment of $H^-$. Third, mutual neutralization, ${\rm
H}^-+{\rm H}^+ \rightarrow 2{\rm H}$ with a rate coefficient
$k_4=7\times 10^{-7} T^{-0.5}$ cm$^3$ s$^{-1}$. Because we are
interested mostly in denser environments (like Milky Way diffuse,
translucent and dense clouds),
where most of the mass is located in clouds with densities of at least
n$_H \sim$ 10$^2$cm$^{-3}$ that
are exposed to modest radiation fields (little star formation), the
disappearance of H$^{-}$ is mainly due to the reaction with H to form
H$_2$. Hence, one can neglect the destruction of H$^-$ by radiation or
through reaction with H$^+$ (Donahue \& Shull 1991). If we consider
gas associated with HII regions that are embedded in galaxies, the
disappearance of
H$^{-}$ is due to the strong radiation of the nearby stars. This
process is important for low ($\sim 1$ cm$^{-3}$) density gas in the local
Universe (z$\le$1) where the star formation rate is higher and the mean
density is lower, but becomes small compared to reactions with H
for gas of higher density at redshifts larger than 2.
Therefore, in the regions that we are
interested in, we can write the equilibrium H$^{-}$ density as
\begin{equation}
n_{\rm H^{-}}=\frac{k_1 n_e}{k_2},
\end{equation}
and we write the H$_2$ formation rate through the H$^-$
route as
\begin{equation}
R_g=k_1 n_{\rm H}^2 \xi_e,
\end{equation}
with $\xi_e$, the electron abundance given by
$\xi_e={{n_e}\over{n_{\rm H}}}$.
Note that in the numerical computations of the cosmological model described
below all destruction routes have been included.

\section{H$_2$ Formation at High Redshift}

\subsection{A Comparison Between the H$^-$ and Dust Grain Routes to H$_2$ Formation}

\begin{figure}
\includegraphics[width=16cm,keepaspectratio,angle=-90]{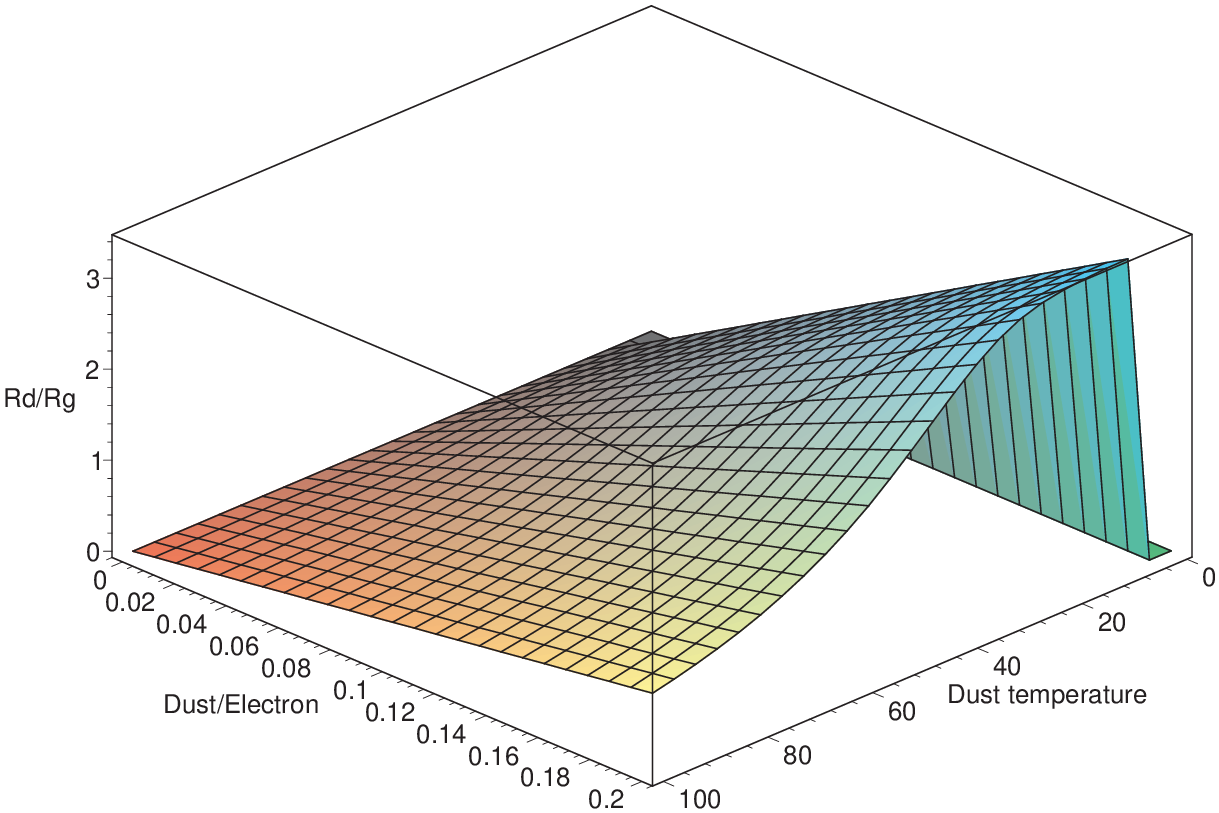}
\caption{Dust to H$^-$ route ratio for the \hm\ formation rate as
a function of $\frac{\xi_d}{\xi_e}$ and the temperature of the dust
$T_d$. The gas temperature is set at 100 K.}
\label{Tg1}
\end{figure}

\begin{figure}
\includegraphics[width=16cm,keepaspectratio,angle=-90]{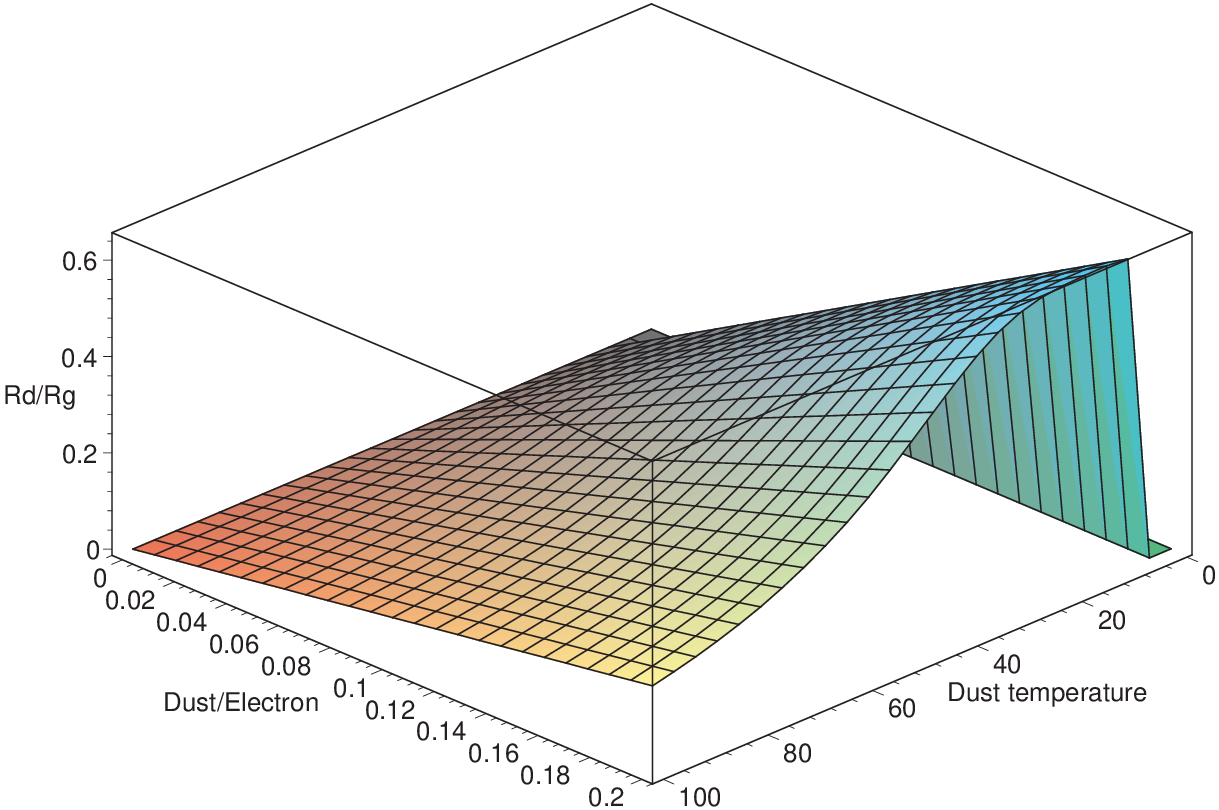}
\caption{Dust to H$^-$ route ratio for the \hm\ formation rate as
a function of $\frac{\xi_d}{\xi_e}$ and the temperature of the dust
$T_d$. The gas temperature is set at 500 K.}
\label{Tg2}
\end{figure}

\begin{figure}
\includegraphics[width=16cm,keepaspectratio,angle=-90]{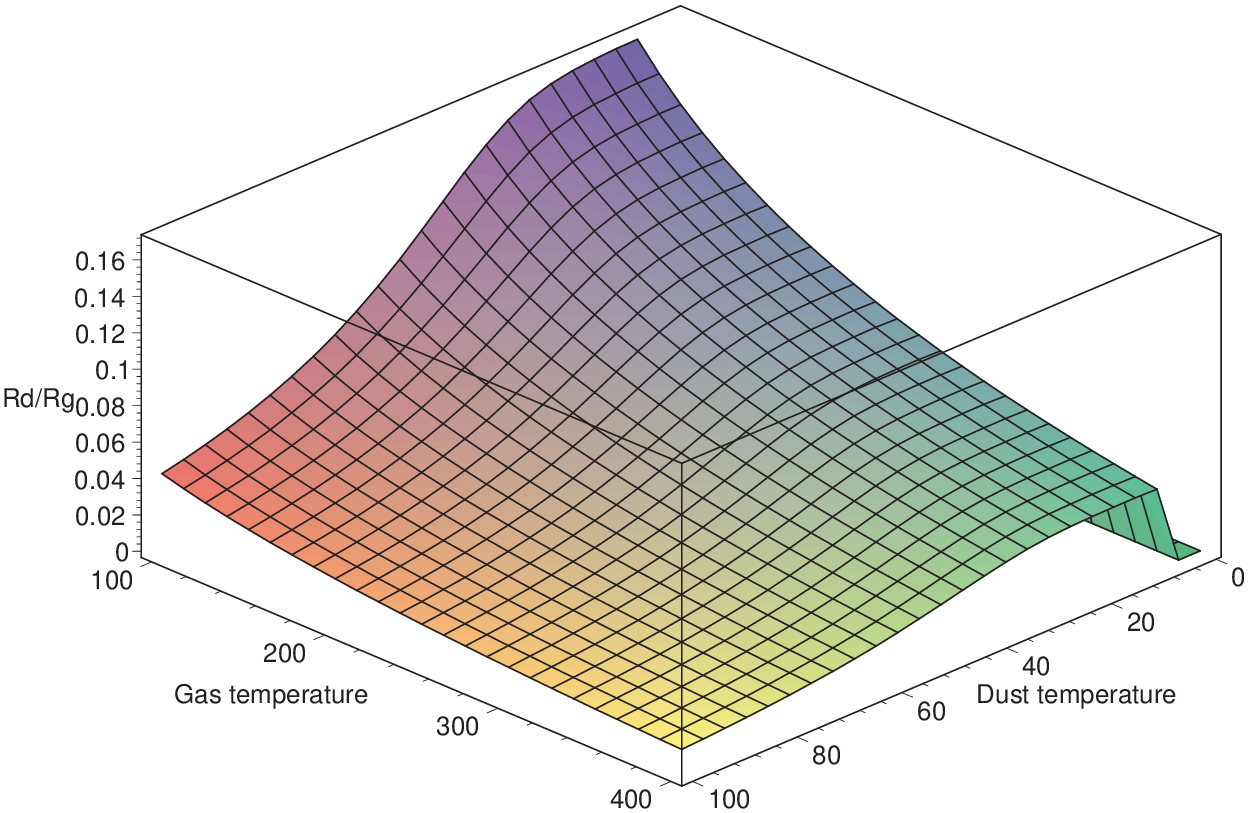}
\caption{Dust to H$^-$ route ratio for the \hm\ formation rate as a
function of the temperature of the dust $T_d$ and of the gas $T_g$.
The dust to electron ratio $\frac{\xi_d}{\xi_e}$ is set at 0.01.}
\label{xi01}
\end{figure}

\begin{figure}
\includegraphics[width=16cm,keepaspectratio,angle=-90]{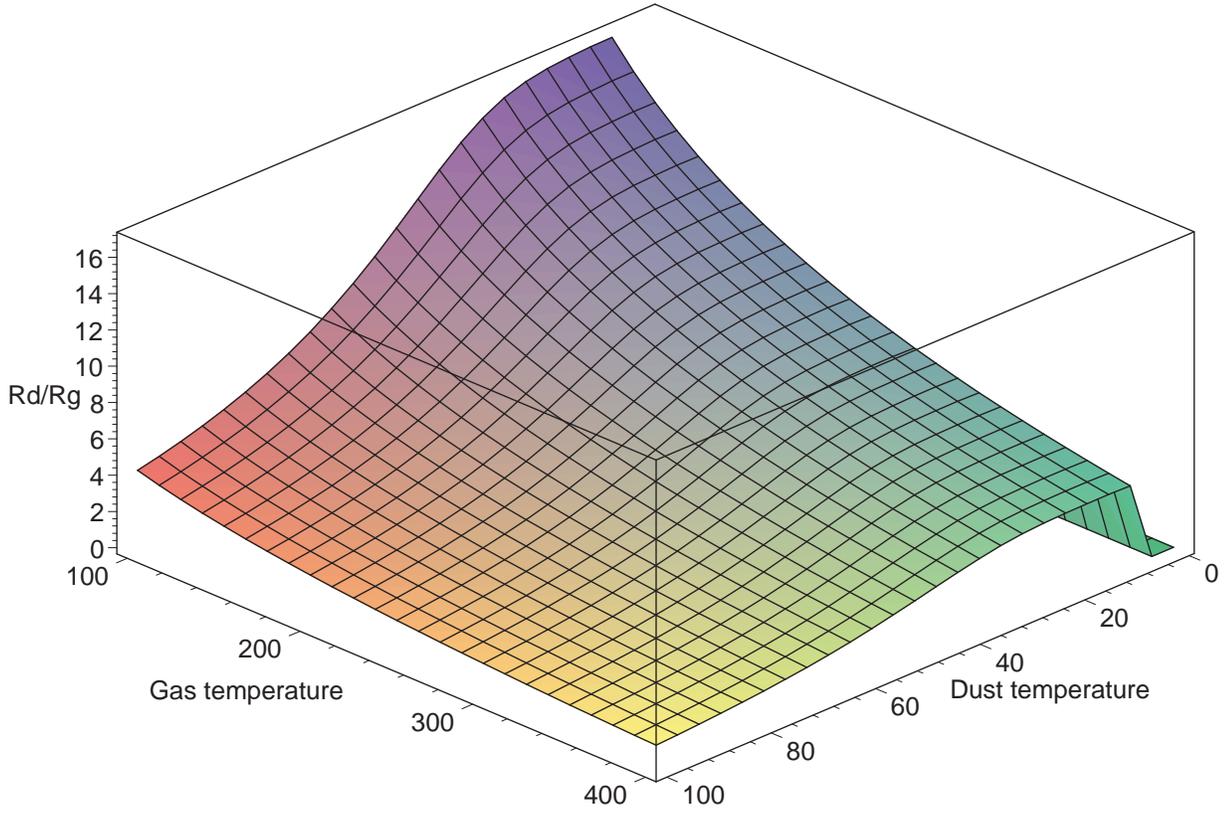}
\caption{Dust to H$^-$ route ratio for the \hm\ formation rate as
a function of the temperature of the dust $T_d$ and of the gas $T_g$.
The dust to electron ratio $\frac{\xi_d}{\xi_e}$ is set at 1.}
\label{xi02}
\end{figure}

\begin{figure}
\includegraphics[width=16cm,keepaspectratio,angle=-90]{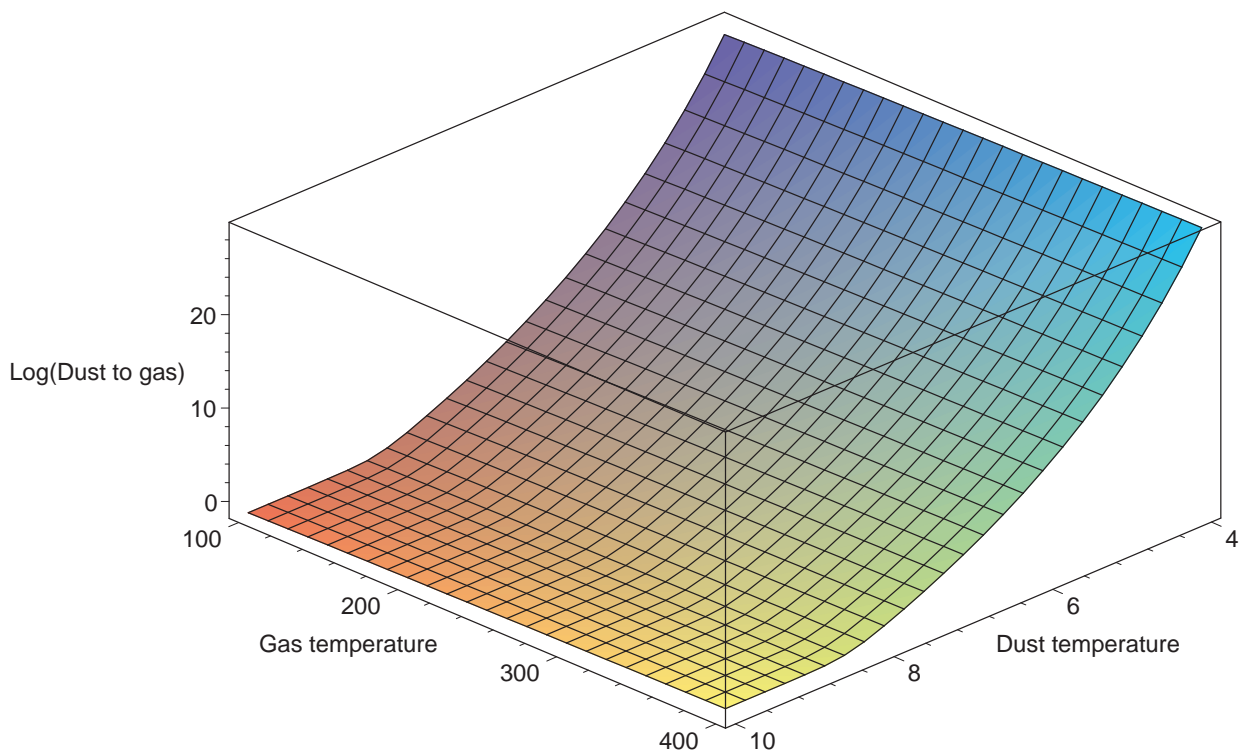}
\caption{This surface represents the different conditions under which
the dust route contribution to the H$_2$ formation rate is equal to
that of the H$^-$ route for low dust temperatures. We plot the dust to
electron ratio $\frac{\xi_d}{\xi_e}$ in log because this parameter
varies a lot with dust temperature for the range considered.
Log($\frac{\xi_d}{\xi_e}$) is plotted as a function of the temperature
of the dust $T_d$ and of the temperature of the gas $T_g$.}
\label{LT}
\end{figure}

\begin{figure}
\includegraphics[width=16cm,keepaspectratio,angle=-90]{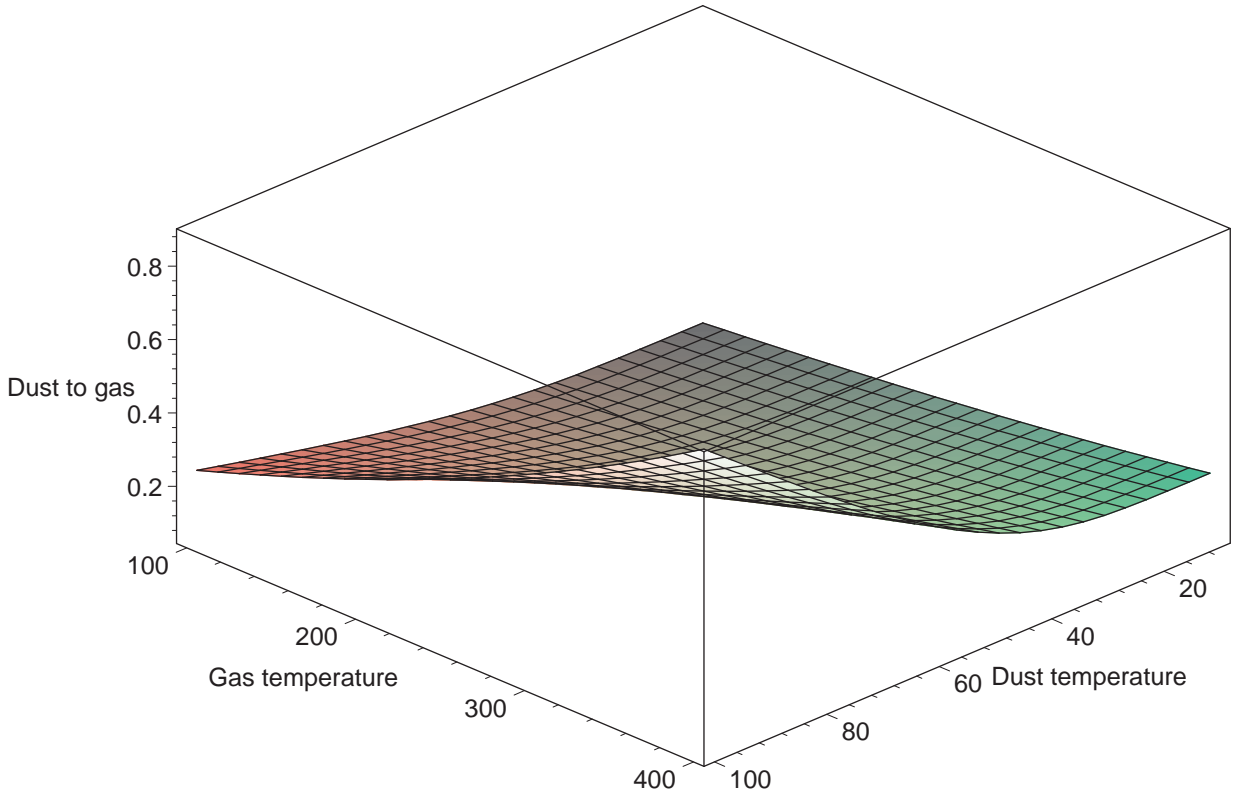}
\caption{This surface represents the different conditions under which
the dust route contribution to the H$_2$ formation rate is equal to
that of the H$^-$ route for high dust temperatures. The dust to
electron ratio $\frac{\xi_d}{\xi_e}$ is plotted as a function of $T_d$
and $T_g$.}
\label{HT}
\end{figure}

We first compare the micro-physics of H$_2$ formation on dust grains
and in the gas phase by treating the most influential variables as
completely free. Our motivation is the substantial uncertainty that
still exists regarding the dust and gas temperature of primordial gas
as well as the free electron abundance and FUV radiation field at high
redshift. The general results of this section can be incorporated by
the reader into any cosmological model. We compute the ratio of H$_2$
formation through the dust and gas routes as
\begin{equation}
\frac{R_d}{R_g}=1.65 \times 10^3 \frac{\xi_d}{\xi_e} \frac{S_{\rm
H}}{{T_g}^{0.8779}}\epsilon_{\rm H_2} \sqrt{\frac{T_g}{100}}.
\end{equation}
Figs.~\ref{Tg1}, ~\ref{Tg2} and ~\ref{xi01} and~\ref{xi02} show the
ratio $R_d/R_g$ as functions of $T_d$, $T_g$, and
$\frac{\xi_d}{\xi_e}$. We find that the dust to electron ratio
$\frac{\xi_d}{\xi_e}$ is the dominant parameter for the behavior of
the ratio $\frac{R_d}{R_g}$. In Figs.~\ref{HT} and ~\ref{LT} we
present the surface $\frac{R_d}{R_g}$=1 as a function of the 3 free
parameters. These two figures show for which conditions the gas phase
route and the dust grain route to \hm\ formation are equal. At low
dust grain temperatures, as shown by Fig.\ ~\ref{LT}, the required
ratio $\frac{\xi_d}{\xi_e}$ varies considerably. This is easily
explained by the fact that at these temperatures, \hm\, once formed,
mostly stays on dust grains (the temperature of the grain is then too
low to enable evaporation). Therefore, for these considered
temperatures, \hm\ formation through the dust route is extremely
small, and the physical conditions to obtain the equality through the
two different routes give a extremely high ratio
$\frac{\xi_d}{\xi_e}$. This range of temperatures is quite small
(between 0 and 10K). Because we know that the grain temperature is
typically larger than 10 K, we consider a range of temperatures between
10 and 100 K for the dust grains. For this considered range, as shown
in Fig.~\ref{HT}, the ratio $\frac{\xi_d}{\xi_e}$ required for
$\frac{R_d}{R_g}$=1, varies slightly with the dust and the gas
temperature. In conclusion, \hm\ formation through the H$^-$ route is
equal to \hm\ formation through the dust route if the ratio
$\frac{\xi_d}{\xi_g}$ lies between 0.1 and 0.8.

\subsection{Cosmological Evolution of Physical Quantities}

\subsubsection{Model}

In order to make a cosmological assessment of the relative importance
of the dust grain and H$^-$ route contributions to the total H$_2$
formation rate, we adopt a cosmological model for the density, dust
abundance, electron abundance and radiation field strength as a
function of redshift. Our microscopic model shows that the equivalence
of the dust grain and H$^-$ route to form H$_2$ occurs for a dust to
electron ratio $\frac{\xi_d}{\xi_e}$ between 0.1 and 0.8. Therefore,
in this section, we construct a cosmological model to estimate at
which redshift this equivalence occurs, if at all. Like in Norman \&
Spaans (1997) and Hirashita, Hunt, \& Ferrara (2002), we consider a
disk galaxy with a radius $R_{disk}$ and a scale height
$H$. The radial size of the galaxy, $R_{disk}$, formed at a certain redshift
is given by
$R=10({{\Omega_{b,g}}\over{0.01}})^{-1/3} (1+z)^{-1}$ kpc, where
$\Omega_{b,g}$ is the baryonic mass fraction in the protogalactic
disk, following the treatment of Kauffmann (1996) for a biasing
parameter of $b\approx 1.5$. The height to disk size ratio, $\eta$ =
H/R, as discussed by Norman \& Spaans (1997), is of the order of
0.01-0.03, although this value is uncertain and can be as high as 0.1
(Hirashita, Hunt, \& Ferrara 2002). We wish to emphasize that the
model galaxy constructed here is intended to represent a dwarf or
sub-$L^*$ disk galaxy, typical of the bulk of all disk galaxies, that
starts to form stars at a redshift of a few and continues to do so at
a relatively vigorous pace (much like the Milky Way and the LMC). We would
like to draw attention to the
work of Glover (2003) that investigates as well the effects of metallicity on
the relative contributions of the gas phase and dust grain formation routes of
H$^2$
formation. Our work aims to explore the cosmological
dependence of the H$_2$ formation rate, the metallicity and the star
formation rate (SFR) for a fiducial galaxy, rather than a range of
different environments, as in Glover (2003). Also, we point out that
our treatment of \hm\ formation on dust grains follows the
semi-classical approach of cazaux \& Tielens (2004) which takes into
account the characteristics of the grain surfaces that are essential to
understand the association of H atoms on a grain, as well as the
mobility of the different atoms under the tunneling effect and thermal
hopping.

\subsubsubsection{Electron Density}

The adopted electron fraction, $\xi_e$, in the optically thick limit,
is described by Kitayama \& Ikeuchi (2000) for the equilibrium between
atomic H formation and ionization:
\begin{equation}
\xi_e = \sqrt{\frac {\Gamma_{HI}}{\alpha_H n_H}},
\end{equation}
where $\Gamma_{HI}$ is the photoionization coefficient of the H atoms,
$\alpha_H$ is the hydrogen recombination rate to the ground level
(Spitzer 1978) written as $\alpha_{\rm H}$=2.50$\times 10^{-10}T_g^{-0.7}$.
\indent We consider the radiation field to be a combination of a
background UV radiation field, an internally generated UV radiation
field and cosmic rays.
\begin{equation}
\Gamma_{HI}=\Gamma_{HI}^{\rm{background}}+\Gamma_{HI}^{\rm{internal}}+\Gamma_{HI}^{\rm{cosmic rays}}.
\end{equation}
In our model, we consider the gas to be optically thick around 1000 \AA
, and the spectral index of the radiation field to be $\alpha=5$,
which is typical for a radiation field produced by massive stars. Note
that a spectral index of $\alpha=1$ would be appropriate for a
radiation field produced by quasars.  In our model we consider only
the contribution made by massive stars. Therefore, the hydrogen ionization
rate, as discussed in Kitayama \& Ikeuchi (2000), can be
written as
\begin{equation}
\Gamma_{HI}^{\rm{background}}=8.18\times 10^{-13} I_{21} \tau_{\perp}^{-\frac{8}{3}}(\nu_{HI}) s^{-1},
\end{equation}
where $\tau_{\perp}$ is the optical depth perpendicular to the disk,
given by $\tau_{\perp}=6.3\times10^{-18}N_{\perp}$ with N$_{\perp}$
the hydrogen column density perpendicular to the disk. I$_{21}$ is the
UV background intensity, in units of 10$^{-21}$ ergs s$^{-1}$
cm$^{-2}$ str$^{-1}$ Hz$^{-1}$, which depends on the redshift, and can
be written (Kitayama \& Ikeuchi 2000) as
\begin{equation}
I_{21}=\left(\frac{1+z}{7}\right)^{-6}\hspace{3cm}    6 \le z \le 20
\end{equation}
\begin{equation}
I_{21}=1		\hspace{3cm}     3 \le z \le 6
\end{equation}
\begin{equation}
I_{21}=\left(\frac{1+z}{4}\right)^{4} \hspace{3cm}  0 \le z \le 3.
\end{equation}
The internal UV radiation field, as discussed by Norman \& Spaans
(1997), results from stellar emission and thus depends on the star
formation rate. Therefore, we can use equation (12) to obtain
\begin{equation}
\Gamma_{HI}^{\rm{internal}}=8.18\times 10^{-13} I_{\rm{internal}} \tau_{\parallel}^{-\frac{8}{3}}(\nu_{HI}) s^{-1}
\end{equation}
where $I_{\rm{internal}}$ has the same units as $I_{21}$ and is
directly proportional to the star formation rate (SFR), and
$\tau_{\parallel}=6.3\times10^{-18}N_{\parallel}$ with N$_{\parallel}$
the hydrogen column density along the radius of the disk, where the disk
is assumed to be exponential. We compute
the SFR in our model using Hirashita \& Ferrara (2002) at high
redshift z$\geq$5, and we match it onto the Madau plot for
z$\leq$5. At low redshift, we assume the SFR to be constant with a
value comparable to the Galactic value of $3 M_\odot$ yr$^{-1}$.  For
redshifts between 3 and 9, we parameterize the SFR as $a(1+z)^b$, with
$a$ and $b$ determined by the limit conditions at $z=3$ and $z=9$. At
high redshift, from 9 to 20, we assume the SFR to be constant with a
value of 0.003 M$_\odot$ yr$^{-1}$. Hence,
\begin{equation}
SFR=0.003  \hspace{3cm} \hspace{0.3cm} 9 \le z \le 20
\end{equation}
\begin{equation}
SFR=10^5 \ (1+z)^{-7.5} \hspace{1.9cm} 3 \le z \le 9
\end{equation}
\begin{equation}
SFR=3  \hspace{3cm}  \hspace{0.9cm}   0 \le z \le 3
\end{equation}
The above is a purely pragmatic approach whose only purpose is to
define a reasonable star formation history for a model disk
galaxy. This parameterization is not intended as an attempt to explain
the Madau plot.

For the hydrogen column densities parallel and perpendicular to the
disk, we consider in our calculations a disk with a radius $R_{disk}$
and an height $H$ as given above, and with a height to disk size ratio
$\eta$. Therefore, the hydrogen column densities are written as
\begin{equation}
N_{\parallel}=n_{\rm H}R_{disk},
\end{equation}
\begin{equation}
N_{\perp}=n_{\rm H}\eta R_{disk},
\end{equation}
where the mean density of a collapsed object, $n_{\rm H}$, evolves
with redshift as $n_{\rm H}=5(\frac{1}{100}\frac{1}{\eta})
({{\Omega_{b,g}}\over{0.01}})(1+z)^3$. Note that the density structure
of the ISM under the influence of metallicity-induced phase
transitions causes clouds to form with a density of $\sim 100$
cm$^{-3}$, while the intercloud medium is typically at a density of
$\sim 1$ cm$^{-3}$, for a fiducial graviational pressure $P\sim
N_{\perp}^2$ of $10^4$ K cm$^{-3}$ at $z=0$ (Norman \& Spaans 1997).

We adopt a mean distance between the, smoothly distributed, stellar
sources and the bulk of hydrogen gas equal to $R_{disk}\times \eta$,
typical for a system where most of the baryonic mass and light are
concentrated within $R_{disk}$. However, the obvious inhomogeneity of
any galactic ISM allows radiation to penetrate much deeper into
gaseous regions. This effect is incorporated in an approximate fashion
by scaling the attenuating column with a redshift independent factor
$\delta\sim 0.1$ (Haiman \& Spaans 1999; Wood \& Loeb 2000). 

Finally, the photo-ionization of H atoms due to cosmic rays can be
scaled with the SFR as
\begin{equation}
\Gamma_{HI}^{\rm{cosmic rays}}=3\times 10^{-17} \frac{SFR}{3},
\end{equation}
where the Galactic cosmic ray ionization rate is used for the prefactor
(Spitzer 1978) and the Milky Way is assumed to form stars at a rate of 3
$M_\odot$/yr.
This contribution is important in regions where the HI column is so large that
UV photons can no longer penetrate.

\subsubsubsection{Dust Abundance}

We can compute the dust-to-gas mass ratio, following Norman \& Spaans
(1997) from:
\begin{equation}
\xi_d={{2}\over{3}}\frac{\xi_l\times y\times \beta}{ H_0 \times
M(z_d)} \int_{z_d}^{20}\frac{SFR}{(1+z)^{\frac{5}{2}}}dz,
\end{equation}
where $\xi_l \sim \frac{1}{3}$ is the fraction of metals locked into
grains, $y\sim 0.02$ is the yield of metals like C and O (Woosley \&
Weaver 1995), $\beta=0.12$ is the fraction of stars formed that will
become supernovae (Yepes et al.\ 1997). Note that these parameters are likely
to depend on redshift (Schneider et al.\ 2002). However, in this investigation
we use the Madau plot as a constraint to bootstrap the metallicity budget.
The combination of the above values and our star formation history satisfies
the observed cosmological metal production rate. The value of $\xi_l$ remains
uncertain and assumes that Galactic dust coagulation is applicable at higher
redshifts as well. The weak dependence on dust surface composition that
we find makes us insensitive to elemental composition effects.
Finally, $H_0=75$ km/s/Mpc is the
Hubble constant and $M\sim 7\times 10^{9}$ $M_{\odot}$ is the gas mass of the
galaxy at redshift 0, typical of a galaxy that has converted a
substantial fraction of its original gas mass into stars, but still
contains a large reservoir of gas as well (like the Milky Way and the
LMC). For these numbers, our model galaxy contains about an order of
magnitude more mass in the form of stars as it does in the form of
atomic hydrogen at $z=0$ (Zwaan et al.\ 1997).  Finally, the integral
in the equation above is the contribution to the metallicity of the
gas by stars that have ended their evolution at a redshift larger than
$z_d$.

\subsubsection{Results}

In our study, we consider two different environments that represent
different phases of the ISM. In the subsection ``global
effects'', we concentrate on clouds in the high density phase with
n$_H$$\sim$100 cm$^{-3}$ at $z=0$, and that are the likely sites of
star formation. In the subsection ``proximity effects'',
we concentrate on low density gas in the vicinity of an HII region with
n$_H$$\sim$1 cm$^{-3}$ at $z=0$. The usual redshift scaling for the
density applies (Norman \& Spaans 1997).

\subsubsubsection{Global Effects}

\begin{figure}
\includegraphics[width=16cm,keepaspectratio,angle=-90]{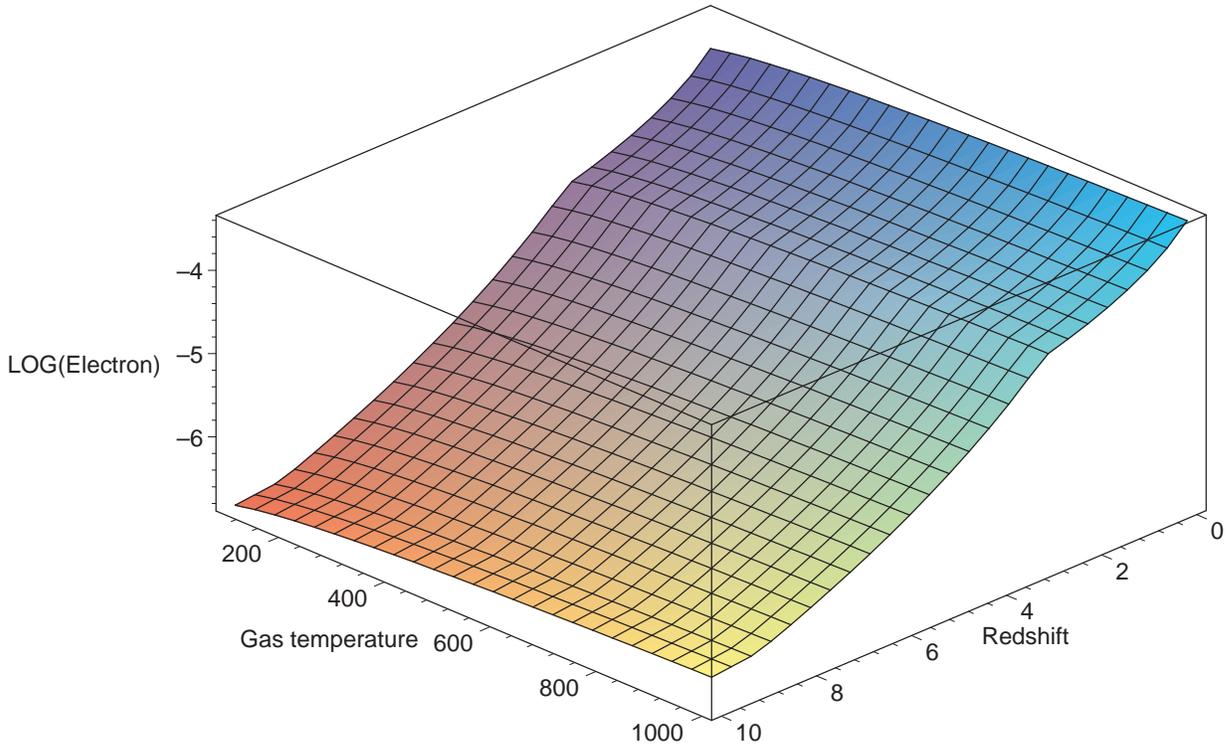}
\caption{The electron density plotted as a function of the temperature of the
gas $T_g$
and the redshift. Note the weak dependence of $\xi_e$ on the gas temperature
in comparison to the redshift dependence.}
\label{xie}
\end{figure}

In Fig.\ ~\ref{xie}, the electron fraction is presented as a function
of the gas temperature and the redshift. We note the strong dependence
of the electron fraction on redshift and the weak dependence it has on
the gas temperature $\sim T^{0.35}$. The decreasing electron fraction
with increasing redshift is easily explained by the fact that
$\tau\sim n_{\rm H} R_{disk}\sim (1+z)^{2}$ and
$J_{21}\sim\tau^{-8/3}$.  The resulting value of $\xi_e$ at $z\sim 0$
is consistent with the electron abundance of the Milky Way and the LMC
in the diffuse/ionized ISM.
 
\begin{figure}
\includegraphics[width=16cm,keepaspectratio,angle=-90]{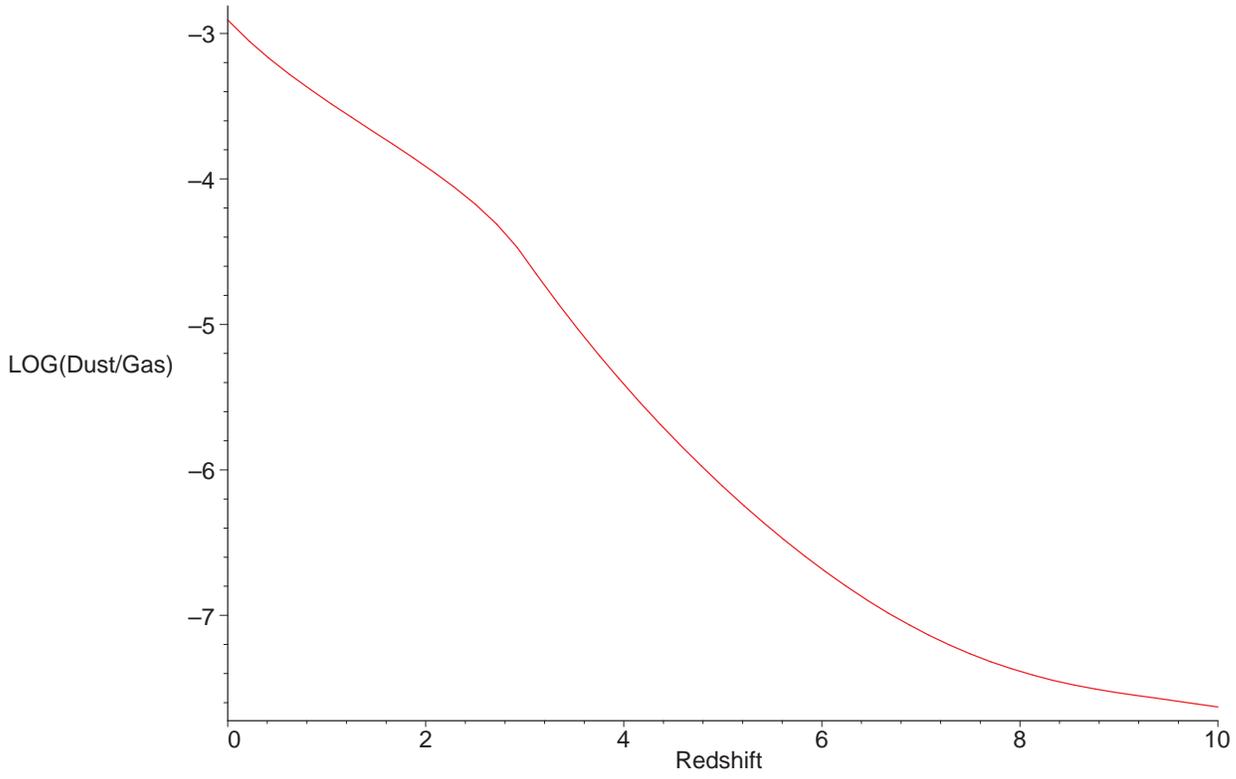}
\caption{The dust-to-gas mass ratio as a function of the redshift for
our model disk galaxy, see the text for parameter values.}
\label{xid}
\end{figure}

In Fig.\ ~\ref{xid} the dust abundance, relative to the
Galactic value $\xi_G$, is presented. The slope of this curve follows
our adopted star formation law. At a redshift of zero the dust
abundance is comparable to that of the LMC. We feel that such a
limiting value for $\xi_d$ is a sensible one to use for the bulk of
the baryonic matter residing in dwarf and sub-L$^*$ galaxies.

\begin{figure}
\includegraphics[width=16cm,keepaspectratio,angle=-90]{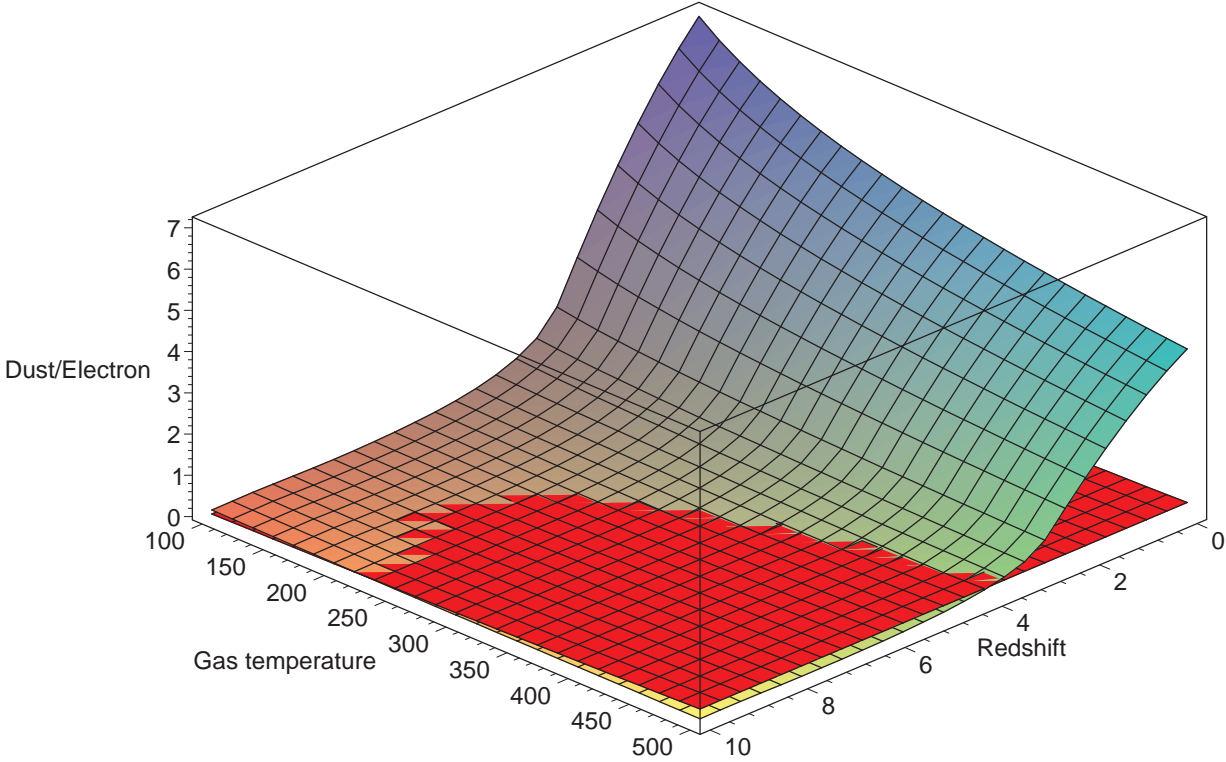}
\caption{The dust to electron ratio $\frac{\xi_d}{\xi_e}$ is plotted
as a function of the temperature of the gas and the redshift, for a
fixed dust temperature of 20 K. The plane surface represents the dust
to electron ratio for which the dust route contribution to the H$_2$
formation rate is equal to that of the H$^-$ route according to our
microscopic model. The other surface represents the cosmological
model. The section of the cosmological surface above the plane surface
determines the cosmological parameters for which the dust route
dominates. Conversely, the section of the cosmological surface below
the plane surface shows the cosmological parameters for which the
H$^-$ route dominates. The figure shows that the dust route can
dominate all the way to $z=10$, for a gas temperature of 100 K, but only
below a redshift of 4, for a gas temperature of 500 K.}
\label{xi}
\end{figure}

\begin{figure}
\includegraphics[width=16cm,keepaspectratio,angle=-90]{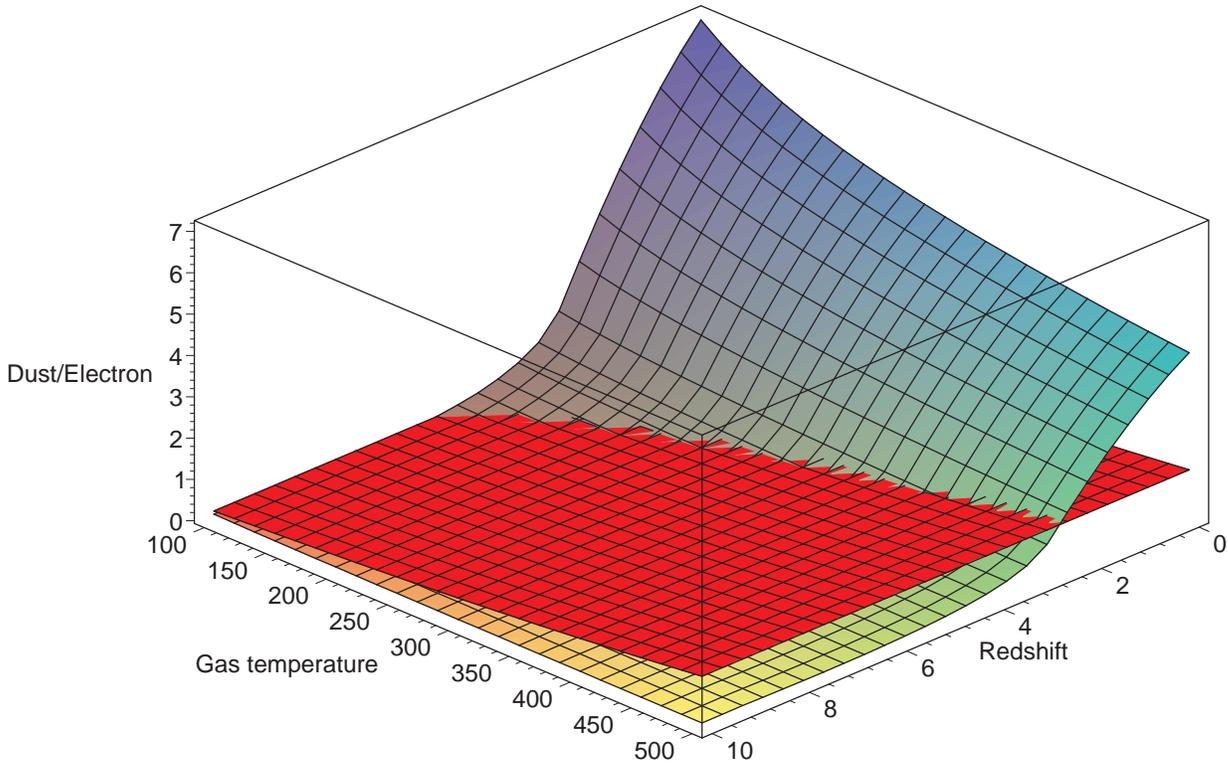}
\caption{Same than figure ~\ref{xi} with our microscopic model (plane
surface) for a fixed dust temperature of 100 K. This figure shows that
the dust route dominates, below a redshift of 6 if the gas temperature
is of the order of 100 K, and below a redshift of 3 if the gas
temperature is of the order of 500 K.}
\label{xi2}
\end{figure}

The ratio of the dust abundance over the electron fraction is
calculated with our cosmological model and is presented in Figs.\
~\ref{xi} and ~\ref{xi2} . In these two figures we overplotted our
microscopic model in order to determine under which conditions the dust
or the H$^-$ route dominates. The flat surface (which is independent
of redshift) represents the dust to electron ratio for which the dust
route contribution to the H$_2$ formation rate is equal to that of the
H$^-$ route according to our microscopic model. The other surface
represents the cosmological model.  The section of the cosmological
surface above the plane surface determines the cosmological parameters
for which the dust route dominates. Conversely, the section of the
cosmological surface below the plane surface shows the cosmological
parameters for which the H$^-$ route dominates. For low dust
temperatures (around 20 K) the dust route is the dominant process at
low redshifts until a redshift of 4, for a gas temperature of 500 K,
and dominates over the entire redshift range considered (z=10-0), for a
gas temperature of 100 K. Note that at z=10,
T$_{dust}$=T$_{CMB}$=30K. For larger dust temperatures (around 100 K),
the dust route dominates from redshift 0 until 6, for a gas
temperature of 100 K. For higher gas temperatures (500K ), the dust
route dominates from redshift 0 until 3.

\subsubsubsection{Proximity Effects}

\begin{figure}
\includegraphics[width=16cm,keepaspectratio,angle=-90]{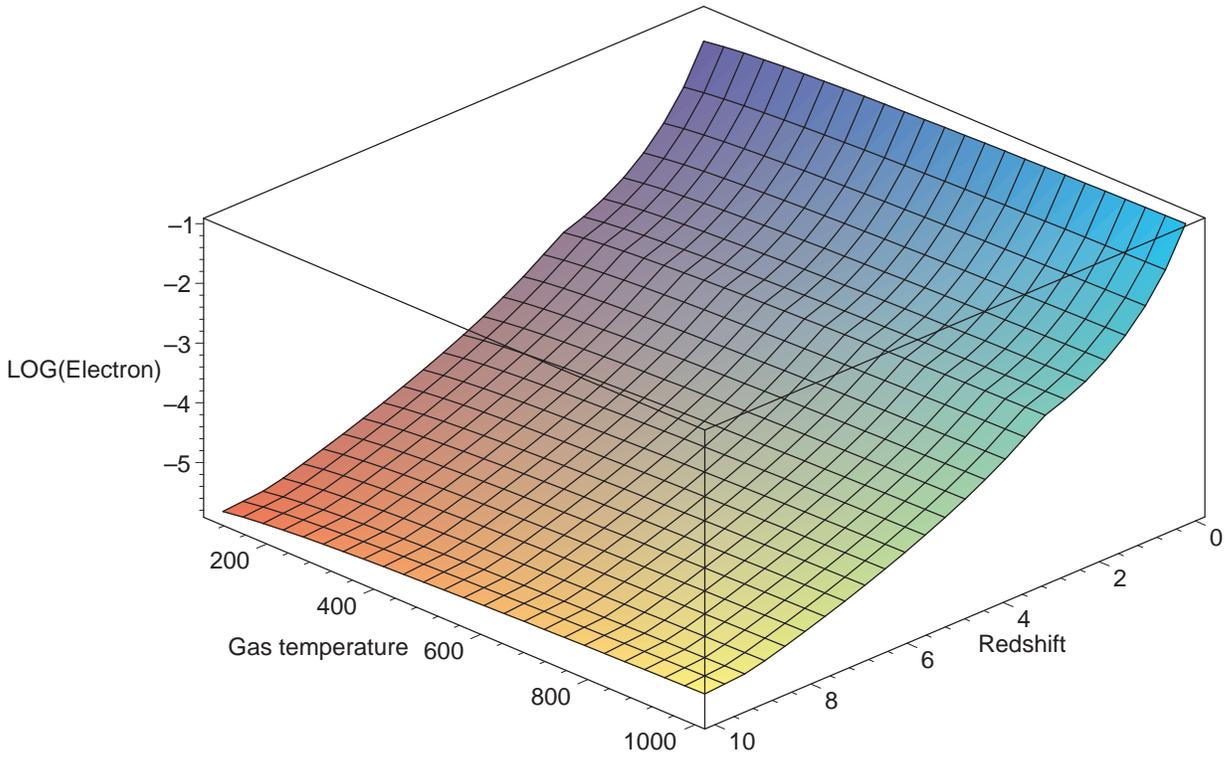}
\caption{The electron density in the low density phase plotted as a
function of the temperature of the gas $T_g$ and the redshift.}
\label{xie2}
\end{figure}

For comparison, we have computed as well a cosmological model with a
density of 1 cm$^{-3}$ and an optical depth parameter of $\tau =10$
at $z=0$,
typical of gas that is exposed to UV photons that have escaped from
nearby HII regions. This region is also assumed to be embedded in a
primordial galaxy whose metallicity and dust abundance evolves as
described in the previous sections. The dust abundance remains the
same than the one calculated for a cloud in a galaxy (see
fig.~\ref{xid}). The electron fraction is higher now since the density and
the optical depth parameters are lower
which allows much more radiation to penetrate and to ionize the
medium. In Fig.\ ~\ref{xie2}, the electron fraction in this region is
presented as a function of the gas temperature and the redshift.
The photo-destruction of H$^-$ is important here and has been included.

\begin{figure}
\includegraphics[width=16cm,keepaspectratio,angle=-90]{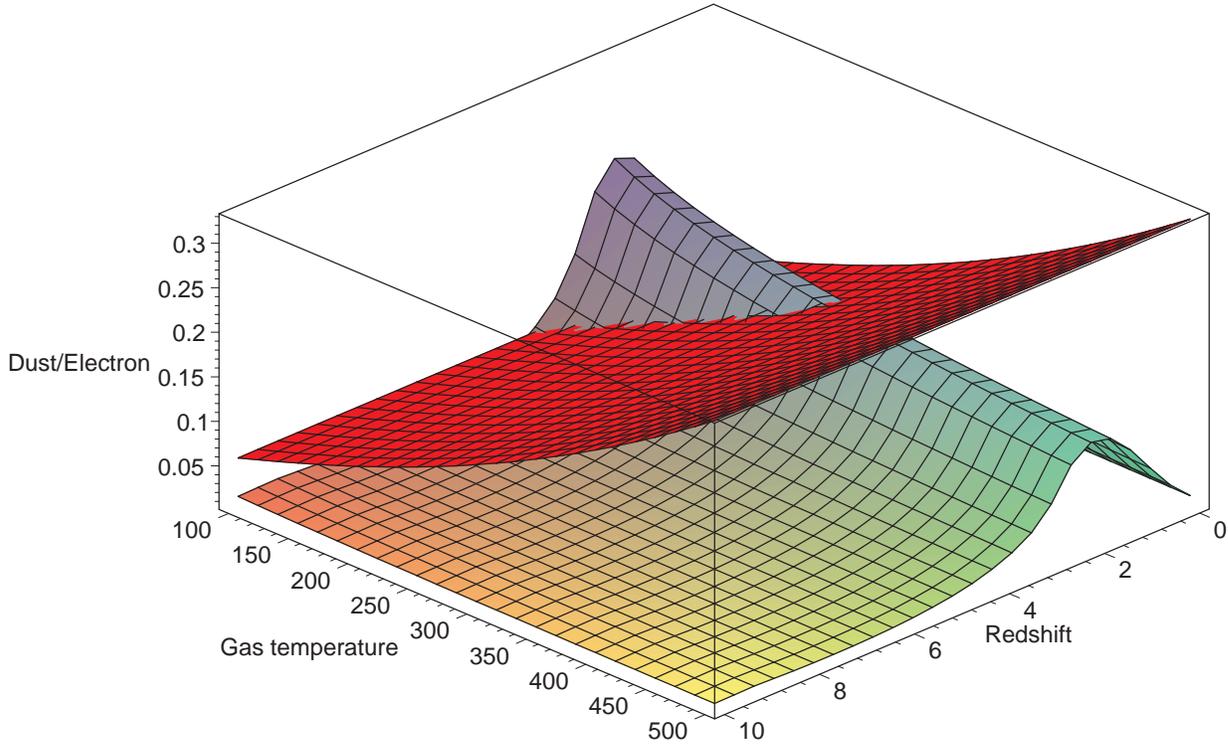}
\caption{Same as figure ~\ref{xi} with a cosmological model describing
low density phase in a primordial galaxy (curved surface) and our
microscopic model for a grain temperature of 20K (plane surface). This
figure shows that in these environments, \hm\ is formed through gas
phase reactions for most gas and dust temperatures.}
\label{xiHII}
\end{figure}

\begin{figure}
\includegraphics[width=16cm,keepaspectratio,angle=-90]{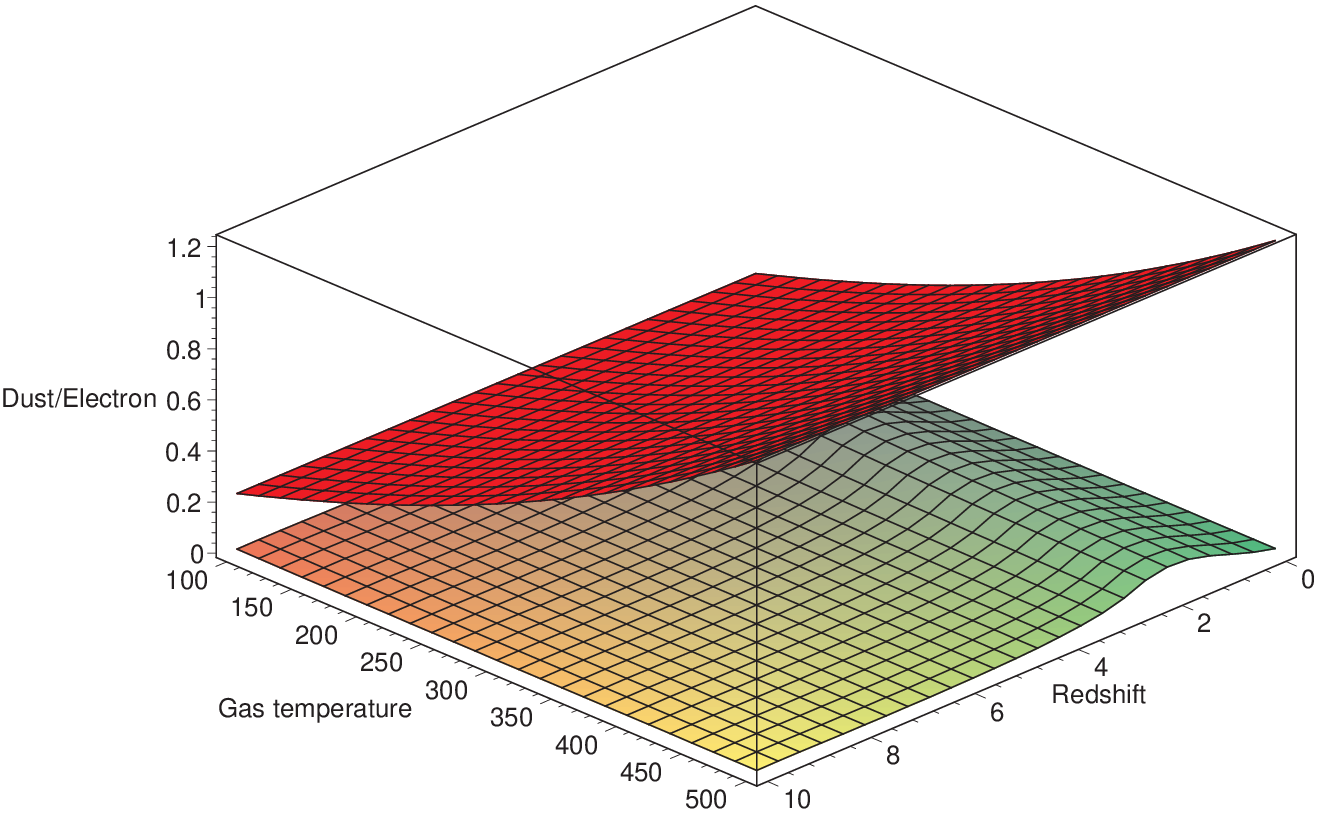}
\caption{Same as figure ~\ref{xi2} with a cosmological model
describing the low density phase in a primordial galaxy (curved surface) and our
microscopic model for a grain temperature of 100 K (plane surface). This
figure shows that in these environments, \hm\ is formed through gas
phase reactions.}
\label{xi2HII}
\end{figure}

It is apparent from Figure ~\ref{xiHII} and ~\ref{xi2HII} that the gas
phase formation route now dominates over the dust surface for most
temperatures. This is caused by the increased electron abundance and
the modest amounts of dust at high redshifts. Still, the bulk of the
gas is generally shielded well and experiences conditions as described
in the ''global effects'' subsection. Hence, proximity effects do not affect
the general trends that we observe.

\section{Conclusions and Discussion}
We have studied the formation of H$_2$ on dust grain surfaces at high
redshift and we have related our results to the contribution made by
gas phase reactions, i.e., through the H$^-$ route. We have found that
the substrate (olivine or amorphous carbon) has a modest impact on the
resulting H$_2$ formation rate. The formation efficiency depends
strongly on the dust temperature below 10 K and above few hundred
kelvins. The role of the gas temperature is more limited, but
suppresses the H$_2$ formation rate above several hundred K due to a
reduced sticking coefficient.  We wish to stress that the
microphysical results on the formation efficiency and the H$_2$
formation rate on dust grains are robust and independent of any galactic
context.

We adopted a cosmological model to determine at which cosmological
parameters the dust and gas phase contributions to the H$_2$ formation
rate are equal, and thus when, in the universe's history, the dust
grain route becomes the dominant H$_2$ formation process. That is,
when the presence of dust, a result of star formation, leads to an
enhancement of the H$_2$ formation rate, which in turn can boost the
H$_2$ abundance (depending on the internal FUV radiation field that
can dissociate H$_2$) and hence the ambient cooling rate by driving a
rapid ion-molecule chemistry that leads to the formation of species
like CO (an excellent low-temperature coolant). Such a cycle
constitutes a positive feedback loop (Hirashita, Hunt, \& Ferrara
2002) and can enhance the star formation rate inside a galaxy.

Our results show that, within the uncertainties of our cosmological
model for the evolution of disk galaxies, the conditions for this
positive feedback can occur at a redshift above 3 which corresponds to
a dust-to-gas mass ratio less than 10$^{-3}$ times the Galactic
value. This redshift range is large and depends strongly on the dust
grain and gas temperatures (microphysics) that we adopted as well as
on the star formation rate (macrophysics) that we used.  Indeed, for
high dust/gas grain temperatures, the atoms desorb/bounce from/off the
grain surface which decreases the \hm\ formation rate. On the other
hand, a high gas temperature favors gas phase formation of H$_2$. This
clearly shows the importance of both dust and gas temperature to
determine which of these two routes dominates. At very high redshift,
$z>15-20$, the temperature of the dust is coupled to the temperature
of the CMB, which increases like $1+z$.  The dust grains are then too
warm to allow an important dust route contribution. So even if dust
would be present at these redshifts, it would not boost the H$_2$
formation rate until the universe had cooled down considerably through
cosmological expansion.

In fact, once the presence of dust boosts the H$_2$ formation rate,
and hence the star formation rate through the enhanced cooling rate,
for redshift higher than $z\sim 3$, the production of stellar photons
will raise the mean dust (and gas) temperature. This constitutes a
minor effect when the shielding dust columns are large and the McKee
criterium is satisfied (McKee 1989), but might be quite important in
the first, metal-poor stages of star formation.  In any case, the
magnitude of the positive feedback that the presence of dust has on
the H$_2$ formation rate requires a careful treatment of the impact
that (enhanced) star formation activity has on the dust (and gas)
thermal equilibrium. We postpone these matters to a future paper.

Finally, recent observations of distant quasars (Bertoldi et al. 2003)
at redshifts around z$\sim$6 showed that these objects possess a
metallicity close to solar. These quasars represent large
over-densities in the Universe and our study concentrates on the
evolution of a typical sub-L* galaxy, that we assume to be more
representative of the average galaxy population. In any case, these
distant quasars, rich in metals, possess both a high dust grain
abundance and the physical conditions to allow efficient \hm\
formation on dust surfaces.

We would like to thank the anonymous referee for his careful reading of the
manuscript and his comments that helped to improve this work.

\clearpage

\begin{table}
\caption{Model parameters for silicate and carbonaceous surfaces.\label{table1}} 
\begin{tabular}{crrrrrrr}
\tableline\tableline\multicolumn{1}{c}{Surface}&
\multicolumn{1}{c}{$E_{H_2}$\tablenotemark{a}} &
\multicolumn{1}{c}{$\mu$\tablenotemark{a}} &
\multicolumn{1}{c}{$E_{S}$\tablenotemark{a}} &
\multicolumn{1}{c}{$E_{{\rm H}_P}$\tablenotemark{a}} &
\multicolumn{1}{c}{$E_{{\rm H}_C}$\tablenotemark{a}} &
\multicolumn{1}{c}{$\nu_{\rm H_2}$\tablenotemark{a}} &
\multicolumn{1}{c}{$\nu_{\rm H_C}$\tablenotemark{a}} \\ \tableline
\multicolumn{1}{c}{unit}&\multicolumn{1}{c}{K} & &
\multicolumn{1}{c}{K} &\multicolumn{1}{c}{K} &\multicolumn{1}{c}{K}&
\multicolumn{1}{c}{s$^{-1}$}& \multicolumn{1}{c}{$s^{-1}$} \\
Olivine&340&0.3&200&650&$\sim$30000&2$\times$10$^{12}$&1$\times$10$^{13}$\\
Carbon&540&0.4&250&800&$\sim$30000&3$\times$10$^{12}$&2$\times$10$^{13}$\\
\tableline
\end{tabular}
\tablenotetext{a}{For more details about the determination and
calculation of these parameters, see Cazaux \& Tielens (2002).}
\tablecomments{ $E_{\rm H_2}$, $E_{{\rm H}_P}$ and $E_{{\rm H}_C}$ are
the desorption energies of H$_2$, physisorbed H (\nhp) and chemisorbed
H (\nhc), and $E_S$ is the energy of the saddle point between two
physisorbed sites. $\mu$ is the fraction of the newly formed \hm\
which stays on the surface and $\nu_{\rm H_2}$ and $\nu_{{\rm H}_C}$
are the vibrational frequencies of \hm\ and H in their surface
sites. The frequency factor for each population $i$ is written as
$\nu_i=\sqrt{\frac{2N_SE_l}{\pi^2 m}}$, where $N_S$ is the surface
number density of sites on the grain, $m$ the mass of the species, and
$E_l$ the energy of the site where the species is bound (physisorbed
of chemisorbed). (a), For more details about the determination and
calculation of these parameters, see Cazaux \& Tielens (2003).}
\end{table}

\clearpage

\clearpage

\clearpage

\clearpage

\end{document}